\documentclass[copyright,creativecommons]{eptcs}
\providecommand{\event}{ACT 2019}
\usepackage{breakurl}
\usepackage{underscore}

\usepackage[T1]{fontenc}
\usepackage[utf8]{inputenc}
\usepackage{amsmath}
\usepackage{cases}
\usepackage{amsthm}
\usepackage{amsfonts}
\usepackage{mathtools}
\usepackage{tikz}
\usepackage{pgf}
\usepackage{graphicx}
\usepackage{subcaption}
\usepackage{multicol}
\usepackage{todonotes}
\usepackage{float}
\usepackage{url}
\usepackage{eurosym}
\usepackage{array}
\usepackage{enumitem}
\usepackage{apxproof}

\usetikzlibrary{arrows,automata}
\usetikzlibrary{positioning,calc,decorations.pathmorphing,decorations.pathreplacing,patterns}

\newtheorem{defi}{Definition}
\newtheoremrep{lemma}{Lemma}
\newtheoremrep{prop}{Proposition}
\newtheoremrep{thm}{Theorem}
\newtheoremrep{coro}{Corollary}

\usepackage{todonotes}

    \renewcommand{\bar}[1]{%
        \overline{#1}}

\newcommand{\cC}[0]{%
\mathcal{C}}

\DeclareMathOperator{\Ob}{Ob}

\DeclareMathOperator{\inj}{inj}

\tikzstyle{copy}=[circle,fill,black,inner sep=2pt]
\tikzstyle{operation}=[circle,draw,black]

\newcolumntype{E}{ >{\centering\arraybackslash} m{.5cm} }
\newcolumntype{F}{ >{\centering\arraybackslash} m{1cm} }
\newcolumntype{G}{ >{\centering\arraybackslash} m{2cm} }
\newcolumntype{H}{ >{\centering\arraybackslash} m{1.6cm}}
\newcolumntype{I}{ >{\centering\arraybackslash} m{3.5cm}}

\newcommand{\executeiffilenewer}[3]{%
\ifnum\pdfstrcmp{\pdffilemoddate{#1}}%
{\pdffilemoddate{#2}}>0%
{\immediate\write18{#3}}\fi%
}

\graphicspath{{diags/}}

\newcommand{\showaxiom}[3]{%
\def\svgscale{0.8}
  \begin{tabular}{>{\centering\arraybackslash} m{#2} E >{\centering\arraybackslash} m{#3}}
\executeiffilenewer{diags/#1-1.svg}{diags/#1-1.pdf}%
{inkscape -z -D --file=diags/#1-1.svg%
--export-pdf=diags/#1-1.pdf --export-latex}%
\input{diags/#1-1.pdf_tex}%
 &
    $=$ &
    \def\svgscale{0.8}
\executeiffilenewer{diags/#1-2.svg}{diags/#1-2.pdf}%
{inkscape -z -D --file=diags/#1-2.svg%
--export-pdf=diags/#1-2.pdf --export-latex}%
\input{diags/#1-2.pdf_tex}%

  \end{tabular}
}

\title{A Complete Language for Faceted Dataflow Programs}
\author{Antonin Delpeuch
\institute{Department of Computer Science\\
University of Oxford}
\email{antonin.delpeuch@cs.ox.ac.uk}}

\def\titlerunning{A Complete Language for Faceted Dataflow Programs}
\def\authorrunning{A. Delpeuch}

\begin{document}

\maketitle

\begin{abstract}
  We present a complete categorical axiomatization of a wide class of
  dataflow programs.  This gives a three-dimensional diagrammatic
  language for workflows, more expressive than the directed acyclic
  graphs generally used for this purpose. This calls for an implementation of
  these representations in data transformation tools.
\end{abstract}

\section*{Introduction}

In the dataflow paradigm, data processing pipelines are built out of
modular components which communicate via some channels. This is a
natural architecture to build concurrent programs and has been studied
in many variants, such as Kahn process
networks~\cite{kahn1974semantics}, Petri
nets~\cite{petri1966communication,kavi1987isomorphisms}, the LUSTRE
language~\cite{halbwachs1991synchronous} or even UNIX processes and
pipes~\cite{walker2009composing}.  Each of these variants comes with
its own requirements on the precise nature of these channels and
operations: for instance, sorting a stream requires the module to read
the entire stream before writing the first value on its output stream,
which violates a requirement called \emph{monotonicity} in Kahn
process networks, but is possible in UNIX.  Categorical accounts of
these process theories have been developed, for instance for Kahn
process networks \cite{stark1991dataflow,hildebrandt2004relational}
or Petri nets~\cite{pratt1991modeling,meseguer1992semantics}.

In this article, we give categorical semantics to programs in
Extract-Transform-Load (ETL) software. These three words refer to the
three main steps of most projects carried out with this sort of
system. Typically, the user extracts data from an existing data source
such as a comma-separated values (CSV) file, transforms it to match a
desired schema (for instance by normalizing values, removing faulty
records, or joining them with other data sources), and loads it into a
more structured information system such as a relational or graph
database. In other words, ETL tools let users move data from one data
model to another. Because the original data source is typically less
structured and not as well curated as the target data store, these
operations are also refered to as data cleansing or wrangling.

ETL tools typically let users manipulate their data via a collection
of operations which can be configured and composed. The way operations
can be composed, as well as the format of the data they act on,
represent the main design choice for these tools: it will determine
what sort of workflow they can represent naturally and efficiently.
We will focus here on the tabular data model popularized the
OpenRefine software~\cite{openrefine}, a widely used open source tool
popular in the linked open data and data journalism
communities.\footnote{See \url{http://openrefine.org/}, we encourage
  viewing the videos or trying the software directly, although this
  article should be readable with no previous knowledge of the tool.}
We give a self-contained description of the tool in
Section~\ref{sec:overview-openrefine}.

We propose a complete categorical axiomatization for this
data model, using two nested monoidal categories.
This gives rise to a three-dimensional diagrammatic language for the workflows,
generalizing the widespread graph-based representation of dataflow pipelines.
The semantics and the complete axiomatization provided makes it possible
to use this model to reason about workflow equivalence using intuitive
graphical rules.

This has very concrete applications: at the time of writing,
OpenRefine has a very limited interface to manipulate workflows, where
the various operations used in the transformation are combined in a
simple list. Graph-based representations of workflows are already
popular in similar tools but are not expressive enough to capture
OpenRefine's model, due to the use of facets, which dynamically change
the route followed by data records in the processing pipeline
depending on their values. Our approach solves this problem by giving
a natural graphical representation which can be understood with no
knowledge of category theory, making it amenable to implementation in
the tool itself.

\section{Categorical semantics of dataflow} \label{sec:dataflow}

Symmetric monoidal categories model an elementary sort of dataflow
pipelines, where the flow is acyclic and deterministic. This is well
known in the applied category theory community: for instance,
\cite{coecke2010quantum} illustrates it by modelling food recipes by
morphisms in such categories.

\begin{defi}
  A \textbf{symmetric monoidal category} (SMC) is a category $\mathcal{C}$
  equipped with a symmetric bifunctor $ \_ \otimes \_ : \cC \times \cC
  \rightarrow \cC$. The tensor product is furthermore required to have
  a unit $I \in \cC$ and to be naturally associative.
\end{defi}

Informally, objects of $\cC$ are stream types and morphisms are dataflow
pipelines binding input streams to output streams. Pipelines
can be composed sequentially, binding the outputs of the first
pipeline to the second, or in parallel, obtaining a pipeline from both
inputs to both outputs. The difference between food and data is that
discarding the latter is not frowned upon: data streams can be discarded and
copied, which makes the category cartesian.

\begin{defi}
  A \textbf{cartesian category} is a symmetric monoidal category $\cC$
  equipped with a natural family of symmetric comonoids $(\delta_A : A
  \to A \otimes A, \bot_A : A \to I)$  such that $\bot_I = 1_I$
  and $\delta_{A \otimes B} = (1_A \otimes s_{A,B} \otimes 1_B) \circ
  (\delta_A \otimes \delta_B)$.  If these conditions are satisfied one
  may write the product as $\times$ instead of $\otimes$.
\end{defi}

\begin{figure}[H]
  \centering
  \begin{subfigure}{0.3\textwidth}
    \centering
\executeiffilenewer{diags/cart-copy.svg}{diags/cart-copy.pdf}%
{inkscape -z -D --file=diags/cart-copy.svg%
--export-pdf=diags/cart-copy.pdf --export-latex}%
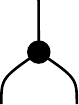%

    \caption{Copying}
  \end{subfigure}
    \begin{subfigure}{0.3\textwidth}
    \centering
\executeiffilenewer{diags/cart-counit.svg}{diags/cart-counit.pdf}%
{inkscape -z -D --file=diags/cart-counit.svg%
--export-pdf=diags/cart-counit.pdf --export-latex}%
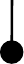%

    \caption{Discarding}
    \end{subfigure}
      \begin{subfigure}{0.3\textwidth}
    \centering
\executeiffilenewer{diags/cart-arbitrary.svg}{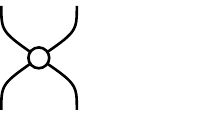}%
{inkscape -z -D --file=diags/cart-arbitrary.svg%
--export-pdf=diags/cart-arbitrary.pdf --export-latex}%
\begingroup%
  \makeatletter%
  \providecommand\color[2][]{%
    \errmessage{(Inkscape) Color is used for the text in Inkscape, but the package 'color.sty' is not loaded}%
    \renewcommand\color[2][]{}%
  }%
  \providecommand\transparent[1]{%
    \errmessage{(Inkscape) Transparency is used (non-zero) for the text in Inkscape, but the package 'transparent.sty' is not loaded}%
    \renewcommand\transparent[1]{}%
  }%
  \providecommand\rotatebox[2]{#2}%
  \newcommand*\fsize{\dimexpr\f@size pt\relax}%
  \newcommand*\lineheight[1]{\fontsize{\fsize}{#1\fsize}\selectfont}%
  \ifx\svgwidth\undefined%
    \setlength{\unitlength}{57.38642826bp}%
    \ifx\svgscale\undefined%
      \relax%
    \else%
      \setlength{\unitlength}{\unitlength * \real{\svgscale}}%
    \fi%
  \else%
    \setlength{\unitlength}{\svgwidth}%
  \fi%
  \global\let\svgwidth\undefined%
  \global\let\svgscale\undefined%
  \makeatother%
  \begin{picture}(1,0.5823432)%
    \lineheight{1}%
    \setlength\tabcolsep{0pt}%
    \put(0,0){\includegraphics[width=\unitlength,page=1]{cart-arbitrary.pdf}}%
    \put(0.32542542,0.29204586){\makebox(0,0)[lt]{\lineheight{1.25}\smash{\begin{tabular}[t]{l}$\alpha$\end{tabular}}}}%
    \put(0.0758664,0.51546518){\makebox(0,0)[lt]{\lineheight{1.25}\smash{\begin{tabular}[t]{l}$\dots$\end{tabular}}}}%
    \put(0.07637789,0.07532711){\makebox(0,0)[lt]{\lineheight{1.25}\smash{\begin{tabular}[t]{l}$\dots$\end{tabular}}}}%
  \end{picture}%
\endgroup%

    \caption{Arbitrary operation}
      \end{subfigure}
      \caption{Generators of a cartesian structure in a SMC}
      \label{fig:cartesian-generators}
\end{figure}

\begin{figure}
  \centering
  \begin{subfigure}{0.45\textwidth}
    \centering
    \begin{tabular}{F E E E F}
      \def\svgscale{0.7}
\executeiffilenewer{diags/cart-right-unit.svg}{diags/cart-right-unit.pdf}%
{inkscape -z -D --file=diags/cart-right-unit.svg%
--export-pdf=diags/cart-right-unit.pdf --export-latex}%
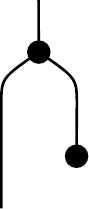%
 &
    $=$ &
          \def\svgscale{0.7}
\executeiffilenewer{diags/cart-identity.svg}{diags/cart-identity.pdf}%
{inkscape -z -D --file=diags/cart-identity.svg%
--export-pdf=diags/cart-identity.pdf --export-latex}%
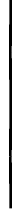%
 &
  $=$ &
        \def\svgscale{0.7}
\executeiffilenewer{diags/cart-left-unit.svg}{diags/cart-left-unit.pdf}%
{inkscape -z -D --file=diags/cart-left-unit.svg%
--export-pdf=diags/cart-left-unit.pdf --export-latex}%
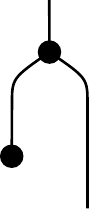%

  \end{tabular}
  \caption{Copying is unital for discarding}
  \end{subfigure}
  \begin{subfigure}{0.45\textwidth}
    \centering
    \begin{tabular}{G E G}
            \def\svgscale{0.7}
\executeiffilenewer{diags/cart-assoc-left.svg}{diags/cart-assoc-left.pdf}%
{inkscape -z -D --file=diags/cart-assoc-left.svg%
--export-pdf=diags/cart-assoc-left.pdf --export-latex}%
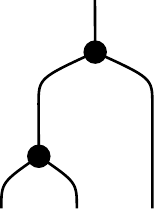%
 &
    $=$ &
          \def\svgscale{0.7}
\executeiffilenewer{diags/cart-assoc-right.svg}{diags/cart-assoc-right.pdf}%
{inkscape -z -D --file=diags/cart-assoc-right.svg%
--export-pdf=diags/cart-assoc-right.pdf --export-latex}%
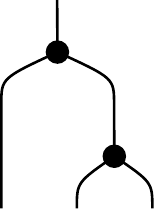%

  \end{tabular}
  \caption{Copying is associative}
  \end{subfigure}

  \begin{subfigure}{0.45\textwidth}
    \centering
    \begin{tabular}{F E F}
            \def\svgscale{0.7}
\executeiffilenewer{diags/cart-counit-arbitrary.svg}{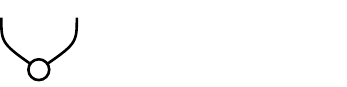}%
{inkscape -z -D --file=diags/cart-counit-arbitrary.svg%
--export-pdf=diags/cart-counit-arbitrary.pdf --export-latex}%
\begingroup%
  \makeatletter%
  \providecommand\color[2][]{%
    \errmessage{(Inkscape) Color is used for the text in Inkscape, but the package 'color.sty' is not loaded}%
    \renewcommand\color[2][]{}%
  }%
  \providecommand\transparent[1]{%
    \errmessage{(Inkscape) Transparency is used (non-zero) for the text in Inkscape, but the package 'transparent.sty' is not loaded}%
    \renewcommand\transparent[1]{}%
  }%
  \providecommand\rotatebox[2]{#2}%
  \newcommand*\fsize{\dimexpr\f@size pt\relax}%
  \newcommand*\lineheight[1]{\fontsize{\fsize}{#1\fsize}\selectfont}%
  \ifx\svgwidth\undefined%
    \setlength{\unitlength}{99.82721605bp}%
    \ifx\svgscale\undefined%
      \relax%
    \else%
      \setlength{\unitlength}{\unitlength * \real{\svgscale}}%
    \fi%
  \else%
    \setlength{\unitlength}{\svgwidth}%
  \fi%
  \global\let\svgwidth\undefined%
  \global\let\svgscale\undefined%
  \makeatother%
  \begin{picture}(1,0.24983928)%
    \lineheight{1}%
    \setlength\tabcolsep{0pt}%
    \put(0,0){\includegraphics[width=\unitlength,page=1]{cart-counit-arbitrary.pdf}}%
    \put(0.18707326,0.04880503){\makebox(0,0)[lt]{\lineheight{1.25}\smash{\begin{tabular}[t]{l}$\alpha$\end{tabular}}}}%
    \put(0.05455294,0.18134202){\makebox(0,0)[lt]{\lineheight{1.25}\smash{\begin{tabular}[t]{l}$\scriptstyle{\dots}$\end{tabular}}}}%
  \end{picture}%
\endgroup%
 &
    $=$ &
    \def\svgscale{0.7}
\executeiffilenewer{diags/cart-counit.svg}{diags/cart-counit.pdf}%
{inkscape -z -D --file=diags/cart-counit.svg%
--export-pdf=diags/cart-counit.pdf --export-latex}%
\input{diags/cart-counit.pdf_tex}%
 $\scriptstyle{\dots}$ \def\svgscale{0.7}%
\executeiffilenewer{diags/cart-counit.svg}{diags/cart-counit.pdf}%
{inkscape -z -D --file=diags/cart-counit.svg%
--export-pdf=diags/cart-counit.pdf --export-latex}%
\input{diags/cart-counit.pdf_tex}%

  \end{tabular}
  \caption{Operations without outputs discard inputs}
    \end{subfigure}
    \begin{subfigure}{0.45\textwidth}
      \centering
      \begin{tabular}{F E F}
                  \def\svgscale{0.7}
\executeiffilenewer{diags/cart-copy-symmetric.svg}{diags/cart-copy-symmetric.pdf}%
{inkscape -z -D --file=diags/cart-copy-symmetric.svg%
--export-pdf=diags/cart-copy-symmetric.pdf --export-latex}%
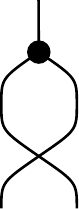%
 &
    $=$ &
              \def\svgscale{0.7}
\executeiffilenewer{diags/cart-copy-long.svg}{diags/cart-copy-long.pdf}%
{inkscape -z -D --file=diags/cart-copy-long.svg%
--export-pdf=diags/cart-copy-long.pdf --export-latex}%
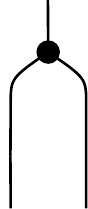%

   \end{tabular}
   \caption{Copying is symmetric}
     \end{subfigure}
   
      \begin{subfigure}{0.4\textwidth}
        \centering
        \begin{tabular}{G E G}
                    \def\svgscale{0.7}
\executeiffilenewer{diags/cart-copying-1.svg}{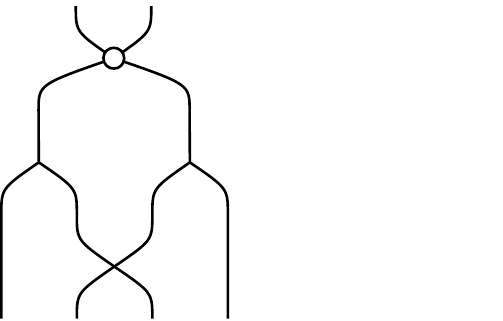}%
{inkscape -z -D --file=diags/cart-copying-1.svg%
--export-pdf=diags/cart-copying-1.pdf --export-latex}%
\begingroup%
  \makeatletter%
  \providecommand\color[2][]{%
    \errmessage{(Inkscape) Color is used for the text in Inkscape, but the package 'color.sty' is not loaded}%
    \renewcommand\color[2][]{}%
  }%
  \providecommand\transparent[1]{%
    \errmessage{(Inkscape) Transparency is used (non-zero) for the text in Inkscape, but the package 'transparent.sty' is not loaded}%
    \renewcommand\transparent[1]{}%
  }%
  \providecommand\rotatebox[2]{#2}%
  \newcommand*\fsize{\dimexpr\f@size pt\relax}%
  \newcommand*\lineheight[1]{\fontsize{\fsize}{#1\fsize}\selectfont}%
  \ifx\svgwidth\undefined%
    \setlength{\unitlength}{144.6653265bp}%
    \ifx\svgscale\undefined%
      \relax%
    \else%
      \setlength{\unitlength}{\unitlength * \real{\svgscale}}%
    \fi%
  \else%
    \setlength{\unitlength}{\svgwidth}%
  \fi%
  \global\let\svgwidth\undefined%
  \global\let\svgscale\undefined%
  \makeatother%
  \begin{picture}(1,0.63833003)%
    \lineheight{1}%
    \setlength\tabcolsep{0pt}%
    \put(0,0){\includegraphics[width=\unitlength,page=1]{cart-copying-1.pdf}}%
    \put(0.27840119,0.52244552){\makebox(0,0)[lt]{\lineheight{1.25}\smash{\begin{tabular}[t]{l}$\alpha$\end{tabular}}}}%
    \put(0,0){\includegraphics[width=\unitlength,page=2]{cart-copying-1.pdf}}%
    \put(0.19075101,0.61180059){\makebox(0,0)[lt]{\lineheight{1.25}\smash{\begin{tabular}[t]{l}$\scriptstyle{\dots}$\end{tabular}}}}%
    \put(0.19101048,0.44189908){\makebox(0,0)[lt]{\lineheight{1.25}\smash{\begin{tabular}[t]{l}$\scriptstyle{\dots}$\end{tabular}}}}%
    \put(0.04561675,0.03367826){\makebox(0,0)[lt]{\lineheight{1.25}\smash{\begin{tabular}[t]{l}$\scriptstyle{\dots}$\end{tabular}}}}%
    \put(0.34758833,0.03367822){\makebox(0,0)[lt]{\lineheight{1.25}\smash{\begin{tabular}[t]{l}$\scriptstyle{\dots}$\end{tabular}}}}%
  \end{picture}%
\endgroup%
 &
  $=$ &
            \def\svgscale{0.7}
\executeiffilenewer{diags/cart-copying-2.svg}{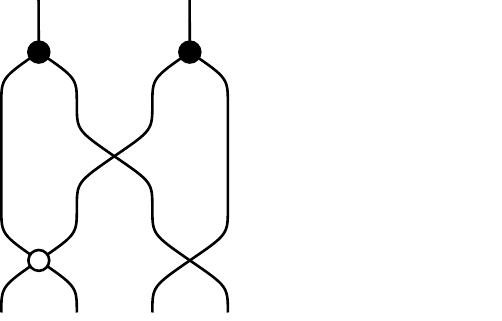}%
{inkscape -z -D --file=diags/cart-copying-2.svg%
--export-pdf=diags/cart-copying-2.pdf --export-latex}%
\begingroup%
  \makeatletter%
  \providecommand\color[2][]{%
    \errmessage{(Inkscape) Color is used for the text in Inkscape, but the package 'color.sty' is not loaded}%
    \renewcommand\color[2][]{}%
  }%
  \providecommand\transparent[1]{%
    \errmessage{(Inkscape) Transparency is used (non-zero) for the text in Inkscape, but the package 'transparent.sty' is not loaded}%
    \renewcommand\transparent[1]{}%
  }%
  \providecommand\rotatebox[2]{#2}%
  \newcommand*\fsize{\dimexpr\f@size pt\relax}%
  \newcommand*\lineheight[1]{\fontsize{\fsize}{#1\fsize}\selectfont}%
  \ifx\svgwidth\undefined%
    \setlength{\unitlength}{144.39566936bp}%
    \ifx\svgscale\undefined%
      \relax%
    \else%
      \setlength{\unitlength}{\unitlength * \real{\svgscale}}%
    \fi%
  \else%
    \setlength{\unitlength}{\svgwidth}%
  \fi%
  \global\let\svgwidth\undefined%
  \global\let\svgscale\undefined%
  \makeatother%
  \begin{picture}(1,0.63850742)%
    \lineheight{1}%
    \setlength\tabcolsep{0pt}%
    \put(0,0){\includegraphics[width=\unitlength,page=1]{cart-copying-2.pdf}}%
    \put(0.12933215,0.11910126){\makebox(0,0)[lt]{\lineheight{1.25}\smash{\begin{tabular}[t]{l}$\alpha$\end{tabular}}}}%
    \put(0,0){\includegraphics[width=\unitlength,page=2]{cart-copying-2.pdf}}%
    \put(0.43058772,0.11910126){\makebox(0,0)[lt]{\lineheight{1.25}\smash{\begin{tabular}[t]{l}$\alpha$\end{tabular}}}}%
    \put(0.03610761,0.0366193){\makebox(0,0)[lt]{\lineheight{1.25}\smash{\begin{tabular}[t]{l}$\scriptstyle{\dots}$\end{tabular}}}}%
    \put(0.34263494,0.03374111){\makebox(0,0)[lt]{\lineheight{1.25}\smash{\begin{tabular}[t]{l}$\scriptstyle{\dots}$\end{tabular}}}}%
    \put(0.03449691,0.20741893){\makebox(0,0)[lt]{\lineheight{1.25}\smash{\begin{tabular}[t]{l}$\scriptstyle{\dots}$\end{tabular}}}}%
    \put(0.34636996,0.20555144){\makebox(0,0)[lt]{\lineheight{1.25}\smash{\begin{tabular}[t]{l}$\scriptstyle{\dots}$\end{tabular}}}}%
    \put(0.19323468,0.61079958){\makebox(0,0)[lt]{\lineheight{1.25}\smash{\begin{tabular}[t]{l}$\scriptstyle{\dots}$\end{tabular}}}}%
  \end{picture}%
\endgroup%

  \end{tabular}
  \caption{Copying the outputs of an operation}
      \end{subfigure}
      \begin{subfigure}{0.4\textwidth}
        \centering
        \begin{tabular}{F E F}
                    \def\svgscale{0.7}
\executeiffilenewer{diags/cart-discarding-1.svg}{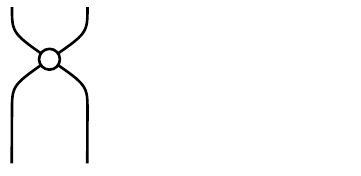}%
{inkscape -z -D --file=diags/cart-discarding-1.svg%
--export-pdf=diags/cart-discarding-1.pdf --export-latex}%
\begingroup%
  \makeatletter%
  \providecommand\color[2][]{%
    \errmessage{(Inkscape) Color is used for the text in Inkscape, but the package 'color.sty' is not loaded}%
    \renewcommand\color[2][]{}%
  }%
  \providecommand\transparent[1]{%
    \errmessage{(Inkscape) Transparency is used (non-zero) for the text in Inkscape, but the package 'transparent.sty' is not loaded}%
    \renewcommand\transparent[1]{}%
  }%
  \providecommand\rotatebox[2]{#2}%
  \newcommand*\fsize{\dimexpr\f@size pt\relax}%
  \newcommand*\lineheight[1]{\fontsize{\fsize}{#1\fsize}\selectfont}%
  \ifx\svgwidth\undefined%
    \setlength{\unitlength}{103.72052908bp}%
    \ifx\svgscale\undefined%
      \relax%
    \else%
      \setlength{\unitlength}{\unitlength * \real{\svgscale}}%
    \fi%
  \else%
    \setlength{\unitlength}{\svgwidth}%
  \fi%
  \global\let\svgwidth\undefined%
  \global\let\svgscale\undefined%
  \makeatother%
  \begin{picture}(1,0.48632461)%
    \lineheight{1}%
    \setlength\tabcolsep{0pt}%
    \put(0,0){\includegraphics[width=\unitlength,page=1]{cart-discarding-1.pdf}}%
    \put(0.20969813,0.32177815){\makebox(0,0)[lt]{\lineheight{1.25}\smash{\begin{tabular}[t]{l}$\alpha$\end{tabular}}}}%
    \put(0,0){\includegraphics[width=\unitlength,page=2]{cart-discarding-1.pdf}}%
    \put(0.08023711,0.44932238){\makebox(0,0)[lt]{\lineheight{1.25}\smash{\begin{tabular}[t]{l}$\scriptstyle{\dots}$\end{tabular}}}}%
    \put(0.09004178,0.19440006){\makebox(0,0)[lt]{\lineheight{1.25}\smash{\begin{tabular}[t]{l}$\scriptstyle{\dots}$\end{tabular}}}}%
  \end{picture}%
\endgroup%
 &
  $=$ &
            \def\svgscale{0.7}
\executeiffilenewer{diags/cart-discarding-2.svg}{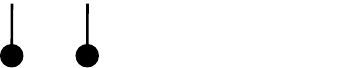}%
{inkscape -z -D --file=diags/cart-discarding-2.svg%
--export-pdf=diags/cart-discarding-2.pdf --export-latex}%
\begingroup%
  \makeatletter%
  \providecommand\color[2][]{%
    \errmessage{(Inkscape) Color is used for the text in Inkscape, but the package 'color.sty' is not loaded}%
    \renewcommand\color[2][]{}%
  }%
  \providecommand\transparent[1]{%
    \errmessage{(Inkscape) Transparency is used (non-zero) for the text in Inkscape, but the package 'transparent.sty' is not loaded}%
    \renewcommand\transparent[1]{}%
  }%
  \providecommand\rotatebox[2]{#2}%
  \newcommand*\fsize{\dimexpr\f@size pt\relax}%
  \newcommand*\lineheight[1]{\fontsize{\fsize}{#1\fsize}\selectfont}%
  \ifx\svgwidth\undefined%
    \setlength{\unitlength}{102.89425707bp}%
    \ifx\svgscale\undefined%
      \relax%
    \else%
      \setlength{\unitlength}{\unitlength * \real{\svgscale}}%
    \fi%
  \else%
    \setlength{\unitlength}{\svgwidth}%
  \fi%
  \global\let\svgwidth\undefined%
  \global\let\svgscale\undefined%
  \makeatother%
  \begin{picture}(1,0.18878503)%
    \lineheight{1}%
    \setlength\tabcolsep{0pt}%
    \put(0,0){\includegraphics[width=\unitlength,page=1]{cart-discarding-2.pdf}}%
    \put(0.08273454,0.15148567){\makebox(0,0)[lt]{\lineheight{1.25}\smash{\begin{tabular}[t]{l}$\scriptstyle{\dots}$\end{tabular}}}}%
  \end{picture}%
\endgroup%

  \end{tabular}
  \caption{Discarding the outputs of an operation}
      \end{subfigure}      

      \caption{Axioms of a cartesian structure in a SMC}
            \label{fig:cartesian}
\end{figure}

The comultiplication $\delta_A$ is the copying map and the counit
$\bot_A$ is the discarding map. One can check that this definition of
cartesian category is equivalent to the usual one, where the product
is defined as the limit of a two-point diagram. The idea behind
defining a cartesian category as a symmetric monoidal category with
extra structure is to obtain a graphical calculus for cartesian
categories. Indeed, morphisms in a SMC can be represented as string
diagrams~\cite{selinger2010survey}. In
Figure~\ref{fig:cartesian-generators} we represent the copying and
discarding maps as explicit operations.\footnote{We draw morphisms with the domain
at the top and the codomain at the bottom.} The equations they satisfy can
then be stated graphically in Figure~\ref{fig:cartesian}.
 
\begin{figure}[H]
  \includegraphics[scale=0.38]{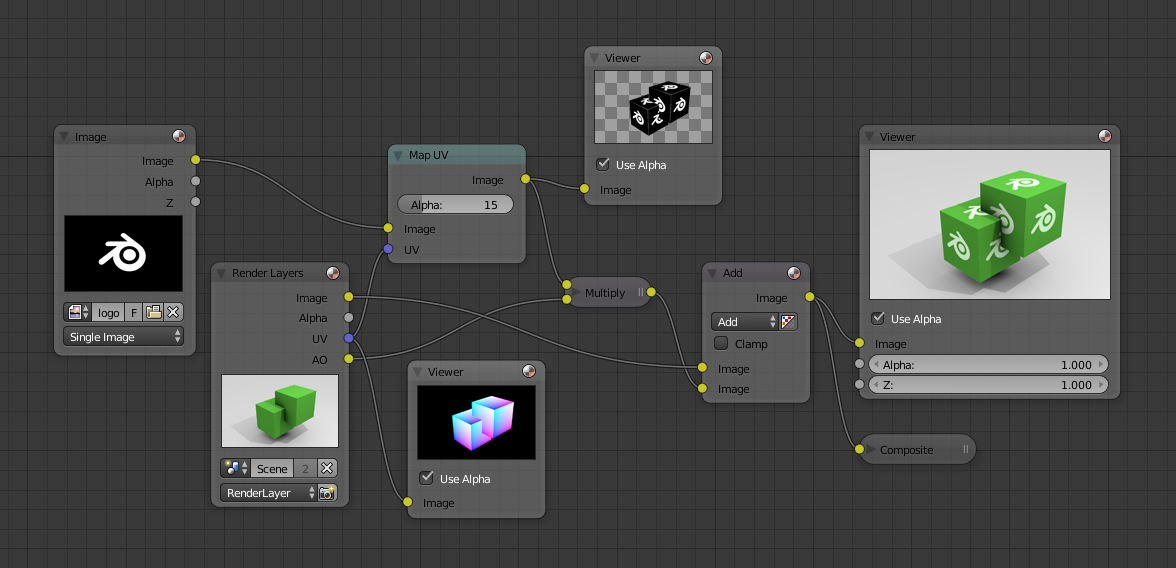}
  \caption{Constructing an image by composing modules in Blender3D. Taken from \url{https://docs.blender.org/manual/en/latest/compositing/introduction.html}, CC BY-SA.}
  \label{fig:blender}
\end{figure}

String diagrams for cartesian categories are essentially directed
acyclic graphs, and this graph-based representation is used in
countless software packages, well beyond ETL tools: for instance,
Figure~\ref{fig:blender} shows a \emph{compositing} workflow in
Blender3D\footnote{\url{https://www.blender.org/}}, where the
graph-based representation of the image transformation pipeline is
manipulated by the user directly.

\section{Overview of OpenRefine} \label{sec:overview-openrefine}

Let us now get into more detail about how OpenRefine works.  Loading a
data source into OpenRefine creates a \emph{project}, which consists
of a simple data table: it is a collection of rows and columns. To
each row and column, a value (possibly null) is associated.

\begin{figure}
  \centering
  \begin{subfigure}{0.8\textwidth}
    \centering
    \begin{tabular}{llr}
      Family name & Given name & Donation \\
      \hline
      Green & Amanda & 25\euro \\
      Dawson & Rupert & 12\euro \\
      de Boer & John & 40\euro \\
      Tusk & Maria & 3\euro \\
    \end{tabular}
    \caption{The initial state of the project}
  \end{subfigure}
  \begin{subfigure}{0.8\textwidth}
    \centering
    \begin{tabular}{lllr}
      Family name & Given name & Full name & Donation \\
      \hline
      Green & Amanda & Amanda Green & 25\euro \\
      Dawson & Rupert & Rupert Dawson & 12\euro \\
      de Boer & John & John de Boer & 40\euro \\
      Tusk & Maria & Maria Tusk & 3\euro \\
    \end{tabular}
    \caption{Applying an operation to create the \texttt{Full name} column}
  \end{subfigure}
    \begin{subfigure}{0.8\textwidth}
    \centering
    \begin{tabular}{lllr}
      Family name & Given name & Full name & Donation \\
      \hline
      GREEN & Amanda & Amanda Green & 25\euro \\
      DAWSON & Rupert & Rupert Dawson & 12\euro \\
      DE BOER & John & John de Boer & 40\euro \\
      TUSK & Maria & Maria Tusk & 3\euro \\
    \end{tabular}
    \caption{Applying an operation to capitalize the \texttt{Family name} column}
  \end{subfigure}

  \begin{subfigure}{0.6\textwidth}
  \centering
  \begin{tikzpicture}
    \node at (0,0) (diag) {%
\executeiffilenewer{diags/simple-OR-workflow.svg}{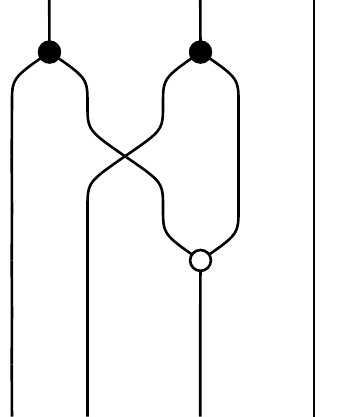}%
{inkscape -z -D --file=diags/simple-OR-workflow.svg%
--export-pdf=diags/simple-OR-workflow.pdf --export-latex}%
\begingroup%
  \makeatletter%
  \providecommand\color[2][]{%
    \errmessage{(Inkscape) Color is used for the text in Inkscape, but the package 'color.sty' is not loaded}%
    \renewcommand\color[2][]{}%
  }%
  \providecommand\transparent[1]{%
    \errmessage{(Inkscape) Transparency is used (non-zero) for the text in Inkscape, but the package 'transparent.sty' is not loaded}%
    \renewcommand\transparent[1]{}%
  }%
  \providecommand\rotatebox[2]{#2}%
  \newcommand*\fsize{\dimexpr\f@size pt\relax}%
  \newcommand*\lineheight[1]{\fontsize{\fsize}{#1\fsize}\selectfont}%
  \ifx\svgwidth\undefined%
    \setlength{\unitlength}{103.96142721bp}%
    \ifx\svgscale\undefined%
      \relax%
    \else%
      \setlength{\unitlength}{\unitlength * \real{\svgscale}}%
    \fi%
  \else%
    \setlength{\unitlength}{\svgwidth}%
  \fi%
  \global\let\svgwidth\undefined%
  \global\let\svgscale\undefined%
  \makeatother%
  \begin{picture}(1,1.15427427)%
    \lineheight{1}%
    \setlength\tabcolsep{0pt}%
    \put(0,0){\includegraphics[width=\unitlength,page=1]{simple-OR-workflow.pdf}}%
    \put(0.62763665,0.43285285){\makebox(0,0)[lt]{\lineheight{1.25}\smash{\begin{tabular}[t]{l}$\alpha$\end{tabular}}}}%
    \put(0,0){\includegraphics[width=\unitlength,page=2]{simple-OR-workflow.pdf}}%
    \put(0.10460612,0.14428428){\makebox(0,0)[lt]{\lineheight{1.25}\smash{\begin{tabular}[t]{l}$\beta$\end{tabular}}}}%
  \end{picture}%
\endgroup%
};
    \node[rotate=45] at (-.5,3.2) {Family name};
    \node[rotate=45] at (1,3.1) {Given name};
    \node[rotate=45] at (2,3) {Donation};
    \node[rotate=45] at (-2.5,-3.1) {Family name};
    \node[rotate=45] at (-1.7,-3) {Given name};
    \node[rotate=45] at (-.4,-2.9) {Full name};
    \node[rotate=45] at (0.8,-2.8) {Donation};
    \draw[dashed] (-2,2) -- (1.7,2);
    \node at (2, 2) {\small{(a)}};
    \draw[dashed] (-2,-1) -- (1.7,-1);
    \node at (2, -1) {\small{(b)}};
    \draw[dashed] (-2,-2) -- (1.7,-2);
    \node at (2,-2) {\small{(c)}};
    
  \end{tikzpicture}
  \caption{The workflow represented as a string diagram. Operation $\alpha$ is concatenation, operation $\beta$ is capitalization}
  \label{fig:example-operation-diagram}
  \end{subfigure}
  
  \caption{Example of an OpenRefine project in its successive states, with the corresponding string diagram}
  \label{fig:example-openrefine-project-states}

\end{figure}

The user can then apply \emph{operations} on this table. Applying an
operation will change the state of the table, usually by performing
the same transformation for each row in the table. Example of
operations include removing a column, reordering columns, normalizing
the case of strings in a column or creating a new column whose values
are obtained by concatenating the values in other columns. Users can
configure these operations with the help of an expression language
which lets them derive the values of a new column from the values in
existing columns.

Unlike spreadsheet software, such expressions are fully evaluated when
stored in the cells that they define: at each stage of the
transformation process, the values in the table are static and will
not be updated further if the values they were derived from change in
the future.  For instance, in the sample project of
Figure~\ref{fig:example-openrefine-project-states}, the first
operation creates a \texttt{Full name} column by concatenating the
\texttt{Given name} and \texttt{Family name} columns. Applying a
second operation to capitalize the \texttt{Family name} column does
not change the values in the \texttt{Full name} column.

Another difference with spreadsheet software, where it is possible to
reference any cell in the expression defining a cell, is that
OpenRefine's expression language only lets the user access values from
the same row. For instance, in the same example project of
Figure~\ref{fig:example-openrefine-project-states}, spreadsheet
software would make it easy to compute the sum of all donations in a
final row. This is not possible in OpenRefine as this would amount to
computing the value of a cell from the value of other cells outside of
its own row.

In other words, operations in OpenRefine are applied row-wise and are
stateless: no state is retained between the processing of rows. It is
therefore simple to parallelize these operations, as they amount to a pure
\emph{map} on the list of rows. This is a simplification: in reality,
there are violations of these requirements (for instance, OpenRefine
offers a sorting operation, and a \emph{records mode} which introduces a restricted form of non-locality). Due to the limited space we do not review these violations here.

\section{Elementary model of OpenRefine workflows} \label{sec:first-model}

So far, OpenRefine fits neatly in the dataflow paradigm presented in
Section~\ref{sec:dataflow}. One can view each column of a
project as a data stream, which can be assigned a type $t \in T$: in
our example project, the first two columns are string-valued and the
third contains monetary values. These data streams are
\emph{synchronous}: the values they contain are aligned to form
rows. An operation $\alpha \in O$ can be seen as reading values from some
columns and writing new columns as output. Because of the
synchronicity requirement, an operation really is just a function from
tuples of input values on the columns it reads to values on the column
it writes.

The schema of a table, which is the list of its column types, can be
naturally represented by the product of the objects representing its
column types. In the example of
Figure~\ref{fig:example-openrefine-project-states}, the initial table
is therefore represented by $S \times S \times M$, where $S$ is the
type of strings and $M$ of monetary values.  Let us call $\alpha : S \times
S \rightarrow S$ the first concatenation operation and $\beta : S
\rightarrow S$ the second capitalization operation.
Figure~\ref{fig:example-operation-diagram} shows a string diagram
which models the workflow of
Figure~\ref{fig:example-openrefine-project-states}.

\begin{defi}
  The category $\mathcal{E}$ of table schema and elementary OpenRefine
  workflows between them is the free cartesian category generated by a
  set of datatypes $D$ as objects and a set of operations $O$ as
  morphisms.
\end{defi}

This modelling of OpenRefine workflows makes it easy to reason about the
information flow in the project. It is possible to rearrange the
operations using the axioms of a cartesian category to show that two
workflows produce the same results. We could add some generating
equations between composites of the generating operations, such as operations
which commute even when executed on the same column for instance.

\begin{figure}
  \centering
    \begin{tabular}{G E >{\centering\arraybackslash} m{5cm}}
\executeiffilenewer{diags/cart-normalization-1.svg}{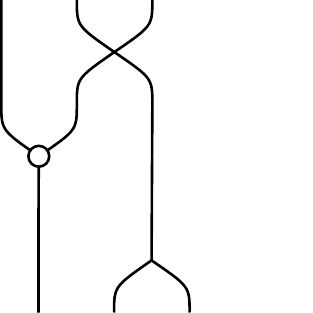}%
{inkscape -z -D --file=diags/cart-normalization-1.svg%
--export-pdf=diags/cart-normalization-1.pdf --export-latex}%
\begingroup%
  \makeatletter%
  \providecommand\color[2][]{%
    \errmessage{(Inkscape) Color is used for the text in Inkscape, but the package 'color.sty' is not loaded}%
    \renewcommand\color[2][]{}%
  }%
  \providecommand\transparent[1]{%
    \errmessage{(Inkscape) Transparency is used (non-zero) for the text in Inkscape, but the package 'transparent.sty' is not loaded}%
    \renewcommand\transparent[1]{}%
  }%
  \providecommand\rotatebox[2]{#2}%
  \newcommand*\fsize{\dimexpr\f@size pt\relax}%
  \newcommand*\lineheight[1]{\fontsize{\fsize}{#1\fsize}\selectfont}%
  \ifx\svgwidth\undefined%
    \setlength{\unitlength}{90.01142826bp}%
    \ifx\svgscale\undefined%
      \relax%
    \else%
      \setlength{\unitlength}{\unitlength * \real{\svgscale}}%
    \fi%
  \else%
    \setlength{\unitlength}{\svgwidth}%
  \fi%
  \global\let\svgwidth\undefined%
  \global\let\svgscale\undefined%
  \makeatother%
  \begin{picture}(1,0.99987304)%
    \lineheight{1}%
    \setlength\tabcolsep{0pt}%
    \put(0,0){\includegraphics[width=\unitlength,page=1]{cart-normalization-1.pdf}}%
    \put(0.20747368,0.49993652){\makebox(0,0)[lt]{\lineheight{1.25}\smash{\begin{tabular}[t]{l}$\beta$\end{tabular}}}}%
    \put(0,0){\includegraphics[width=\unitlength,page=2]{cart-normalization-1.pdf}}%
    \put(0.56992766,0.49993652){\makebox(0,0)[lt]{\lineheight{1.25}\smash{\begin{tabular}[t]{l}$\alpha$\end{tabular}}}}%
    \put(0,0){\includegraphics[width=\unitlength,page=3]{cart-normalization-1.pdf}}%
  \end{picture}%
\endgroup%
 &
  $=$ &
  \begin{tikzpicture}
    \node at (0,0) {%
\executeiffilenewer{diags/cart-normalization-2.svg}{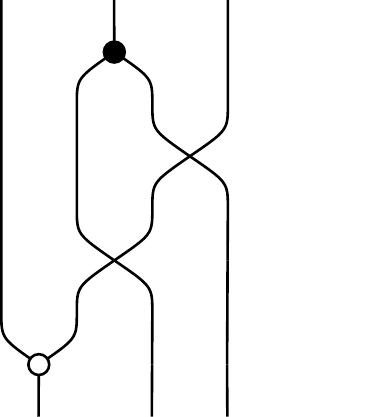}%
{inkscape -z -D --file=diags/cart-normalization-2.svg%
--export-pdf=diags/cart-normalization-2.pdf --export-latex}%
\begingroup%
  \makeatletter%
  \providecommand\color[2][]{%
    \errmessage{(Inkscape) Color is used for the text in Inkscape, but the package 'color.sty' is not loaded}%
    \renewcommand\color[2][]{}%
  }%
  \providecommand\transparent[1]{%
    \errmessage{(Inkscape) Transparency is used (non-zero) for the text in Inkscape, but the package 'transparent.sty' is not loaded}%
    \renewcommand\transparent[1]{}%
  }%
  \providecommand\rotatebox[2]{#2}%
  \newcommand*\fsize{\dimexpr\f@size pt\relax}%
  \newcommand*\lineheight[1]{\fontsize{\fsize}{#1\fsize}\selectfont}%
  \ifx\svgwidth\undefined%
    \setlength{\unitlength}{111.61142483bp}%
    \ifx\svgscale\undefined%
      \relax%
    \else%
      \setlength{\unitlength}{\unitlength * \real{\svgscale}}%
    \fi%
  \else%
    \setlength{\unitlength}{\svgwidth}%
  \fi%
  \global\let\svgwidth\undefined%
  \global\let\svgscale\undefined%
  \makeatother%
  \begin{picture}(1,1.07515875)%
    \lineheight{1}%
    \setlength\tabcolsep{0pt}%
    \put(0,0){\includegraphics[width=\unitlength,page=1]{cart-normalization-2.pdf}}%
    \put(0.1673216,0.13439484){\makebox(0,0)[lt]{\lineheight{1.25}\smash{\begin{tabular}[t]{l}$\beta$\end{tabular}}}}%
    \put(0,0){\includegraphics[width=\unitlength,page=2]{cart-normalization-2.pdf}}%
    \put(0.45963039,0.13439484){\makebox(0,0)[lt]{\lineheight{1.25}\smash{\begin{tabular}[t]{l}$\alpha$\end{tabular}}}}%
    \put(0,0){\includegraphics[width=\unitlength,page=3]{cart-normalization-2.pdf}}%
    \put(0.65315893,0.13439484){\makebox(0,0)[lt]{\lineheight{1.25}\smash{\begin{tabular}[t]{l}$\alpha$\end{tabular}}}}%
  \end{picture}%
\endgroup%
};
    \draw[dashed] (-2.2,1) -- (0.7,1);
    \draw[dashed] (-2.2,-1) -- (0.7,-1);
    \node at (2, -1.5) {operations};
    \node at (2,0) {exchanges};
    \node at (3,1.5) {copying and discarding};
  \end{tikzpicture}
  \end{tabular}
    \caption{A diagram in $\mathcal{E}$ and its normal form}
    \label{fig:cartesian-nf}
\end{figure}

Without loss of generality, we can assume that the generating
operations all have a single generating datatype as codomain, as the
cartesian structure makes it possible to represent generic operations
as composites of their projections. Under these conditions, morphisms
of $\mathcal{E}$ can be rewritten to a normal form, illustrated in Figure~\ref{fig:cartesian-nf}.

\begin{lemma}
  Any morphism $m \in \mathcal{E}$ can be written as a vertical composite
  of three layers: the first one only contains copying and
  discarding morphisms, the second only symmetries and the third only generating
  operations (identities are allowed at each level).
\end{lemma}

All three slices in the decomposition above can be further normalized:
for instance, the cartesian slice can be expressed in left-associative
form, the exchange slice is determined by the permutation it
represents and the operation slice can be expressed in right normal
form~\cite{delpeuch2018normalization}. This gives a simple way to
decide the equality of diagrams in $\mathcal{E}$. Of course, deciding
equality in a free cartesian category just amounts to comparing tuples
of terms in universal algebra. We are only formulating it as a graphical
rewriting procedure to lay down the methodoly for the next section.

\section{Model of OpenRefine workflows with facets}

One key functionality of OpenRefine that we have ignored so far
is its \emph{facets}. A facet on a
column gives a summary of the value distribution in this column. For
instance, a facet on a column containing strings will display the
distinct strings occurring in the column and their number of
occurences. A numerical facet will display a histogram, a scatterplot
facet will display points in the plane, and so on.

Beyond the use of facets to analyze distributions of values, it is
also possible to select particular values in the facet, which selects
the rows where these values are found. It is then possible to run
operations on these filtered rows only. So far our operations ran
on all rows indiscriminately, so we need to extend our model 
to represent operations applied to a filtered set of rows.

We assume from now on a set $F$ of filters in addition to our set of
operations $O$. Each filter $f \in F$ is associated with an object
$T_f \in \mathcal{E}$, the type of data that it filters on. Each
filter can be thought of as a boolean expression that can be evaluated
for each value $v \in T_f$, determining if the value is included or
excluded by the filter. The type $T_f$ is not required to be atomic:
for instance, in the case of a scatterplot filter, two numerical
columns are read.

\begin{figure}
  \centering
  \begin{subfigure}{0.45\textwidth}
    \centering
\executeiffilenewer{diags/operation-def.svg}{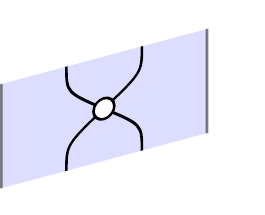}%
{inkscape -z -D --file=diags/operation-def.svg%
--export-pdf=diags/operation-def.pdf --export-latex}%
\begingroup%
  \makeatletter%
  \providecommand\color[2][]{%
    \errmessage{(Inkscape) Color is used for the text in Inkscape, but the package 'color.sty' is not loaded}%
    \renewcommand\color[2][]{}%
  }%
  \providecommand\transparent[1]{%
    \errmessage{(Inkscape) Transparency is used (non-zero) for the text in Inkscape, but the package 'transparent.sty' is not loaded}%
    \renewcommand\transparent[1]{}%
  }%
  \providecommand\rotatebox[2]{#2}%
  \newcommand*\fsize{\dimexpr\f@size pt\relax}%
  \newcommand*\lineheight[1]{\fontsize{\fsize}{#1\fsize}\selectfont}%
  \ifx\svgwidth\undefined%
    \setlength{\unitlength}{76.13642693bp}%
    \ifx\svgscale\undefined%
      \relax%
    \else%
      \setlength{\unitlength}{\unitlength * \real{\svgscale}}%
    \fi%
  \else%
    \setlength{\unitlength}{\svgwidth}%
  \fi%
  \global\let\svgwidth\undefined%
  \global\let\svgscale\undefined%
  \makeatother%
  \begin{picture}(1,0.8118477)%
    \lineheight{1}%
    \setlength\tabcolsep{0pt}%
    \put(0,0){\includegraphics[width=\unitlength,page=1]{operation-def.pdf}}%
    \put(0.49155185,0.4272706){\rotatebox{15}{\makebox(0,0)[lt]{\lineheight{1.25}\smash{\begin{tabular}[t]{l}$\alpha$\end{tabular}}}}}%
    \put(0.20194013,0.58608709){\rotatebox{15}{\makebox(0,0)[lt]{\lineheight{1.25}\smash{\begin{tabular}[t]{l}$A$\end{tabular}}}}}%
    \put(0.48761153,0.66263251){\rotatebox{15}{\makebox(0,0)[lt]{\lineheight{1.25}\smash{\begin{tabular}[t]{l}$B$\end{tabular}}}}}%
    \put(0.20194013,0.01474428){\rotatebox{15}{\makebox(0,0)[lt]{\lineheight{1.25}\smash{\begin{tabular}[t]{l}$C$\end{tabular}}}}}%
    \put(0.48761153,0.09128971){\rotatebox{15}{\makebox(0,0)[lt]{\lineheight{1.25}\smash{\begin{tabular}[t]{l}$D$\end{tabular}}}}}%
  \end{picture}%
\endgroup%

    \caption{Operation $[\alpha] : [A \times B] \to [C \times D]$}
  \end{subfigure}
  \begin{subfigure}{0.45\textwidth}
    \centering
\executeiffilenewer{diags/facet-def.svg}{diags/facet-def.pdf}%
{inkscape -z -D --file=diags/facet-def.svg%
--export-pdf=diags/facet-def.pdf --export-latex}%
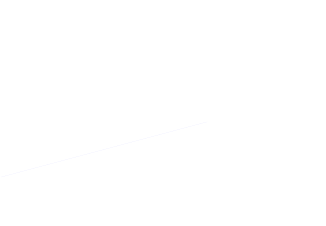%

    \caption{Filter $[f \times U] : [T_f \times U] \rightarrow [T_f \times U, T_f \times U]$}
  \end{subfigure}
  \begin{subfigure}{0.45\textwidth}
    \centering
\executeiffilenewer{diags/join-def.svg}{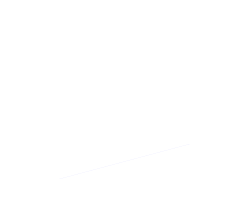}%
{inkscape -z -D --file=diags/join-def.svg%
--export-pdf=diags/join-def.pdf --export-latex}%
\begingroup%
  \makeatletter%
  \providecommand\color[2][]{%
    \errmessage{(Inkscape) Color is used for the text in Inkscape, but the package 'color.sty' is not loaded}%
    \renewcommand\color[2][]{}%
  }%
  \providecommand\transparent[1]{%
    \errmessage{(Inkscape) Transparency is used (non-zero) for the text in Inkscape, but the package 'transparent.sty' is not loaded}%
    \renewcommand\transparent[1]{}%
  }%
  \providecommand\rotatebox[2]{#2}%
  \newcommand*\fsize{\dimexpr\f@size pt\relax}%
  \newcommand*\lineheight[1]{\fontsize{\fsize}{#1\fsize}\selectfont}%
  \ifx\svgwidth\undefined%
    \setlength{\unitlength}{72.00000276bp}%
    \ifx\svgscale\undefined%
      \relax%
    \else%
      \setlength{\unitlength}{\unitlength * \real{\svgscale}}%
    \fi%
  \else%
    \setlength{\unitlength}{\svgwidth}%
  \fi%
  \global\let\svgwidth\undefined%
  \global\let\svgscale\undefined%
  \makeatother%
  \begin{picture}(1,0.81593583)%
    \lineheight{1}%
    \setlength\tabcolsep{0pt}%
    \put(0,0){\includegraphics[width=\unitlength,page=1]{join-def.pdf}}%
    \put(0.46666665,0.01559135){\rotatebox{15}{\makebox(0,0)[lt]{\lineheight{1.25}\smash{\begin{tabular}[t]{l}$T$\end{tabular}}}}}%
    \put(0,0){\includegraphics[width=\unitlength,page=2]{join-def.pdf}}%
    \put(0.21354166,0.66063405){\rotatebox{15}{\makebox(0,0)[lt]{\lineheight{1.25}\smash{\begin{tabular}[t]{l}$T$\end{tabular}}}}}%
    \put(0,0){\includegraphics[width=\unitlength,page=3]{join-def.pdf}}%
    \put(0.68229164,0.56688405){\rotatebox{15}{\makebox(0,0)[lt]{\lineheight{1.25}\smash{\begin{tabular}[t]{l}$T$\end{tabular}}}}}%
  \end{picture}%
\endgroup%

    \caption{Union $J_T : [T, T] \rightarrow [T]$}
  \end{subfigure}
  \begin{subfigure}{0.45\textwidth}
    \centering
\executeiffilenewer{diags/empty-def.svg}{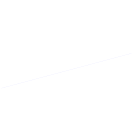}%
{inkscape -z -D --file=diags/empty-def.svg%
--export-pdf=diags/empty-def.pdf --export-latex}%
\begingroup%
  \makeatletter%
  \providecommand\color[2][]{%
    \errmessage{(Inkscape) Color is used for the text in Inkscape, but the package 'color.sty' is not loaded}%
    \renewcommand\color[2][]{}%
  }%
  \providecommand\transparent[1]{%
    \errmessage{(Inkscape) Transparency is used (non-zero) for the text in Inkscape, but the package 'transparent.sty' is not loaded}%
    \renewcommand\transparent[1]{}%
  }%
  \providecommand\rotatebox[2]{#2}%
  \newcommand*\fsize{\dimexpr\f@size pt\relax}%
  \newcommand*\lineheight[1]{\fontsize{\fsize}{#1\fsize}\selectfont}%
  \ifx\svgwidth\undefined%
    \setlength{\unitlength}{38.25bp}%
    \ifx\svgscale\undefined%
      \relax%
    \else%
      \setlength{\unitlength}{\unitlength * \real{\svgscale}}%
    \fi%
  \else%
    \setlength{\unitlength}{\svgwidth}%
  \fi%
  \global\let\svgwidth\undefined%
  \global\let\svgscale\undefined%
  \makeatother%
  \begin{picture}(1,0.85259431)%
    \lineheight{1}%
    \setlength\tabcolsep{0pt}%
    \put(0,0){\includegraphics[width=\unitlength,page=1]{empty-def.pdf}}%
    \put(0.44117647,0.02934842){\rotatebox{15}{\makebox(0,0)[lt]{\lineheight{1.25}\smash{\begin{tabular}[t]{l}$T$\end{tabular}}}}}%
    \put(0,0){\includegraphics[width=\unitlength,page=2]{empty-def.pdf}}%
  \end{picture}%
\endgroup%

    \caption{Empty table: $E_T : [] \rightarrow [T]$}
  \end{subfigure}
  \caption{Generators of $\mathcal{F}$}
  \label{fig:3d-notations}
\end{figure}

\begin{figure}
  \centering
  \begin{subfigure}{0.45\textwidth}
    \centering
    \showaxiom{facet-elim}{3cm}{2cm}
    \caption{Merging a filter immediately does nothing}
    \label{eq:filter-elim}
  \end{subfigure}
  \begin{subfigure}{0.45\textwidth}
    \centering
    \showaxiom{facet-comm}{3cm}{2.5cm}
    \caption{Filters commute even with a common column}
    \label{eq:filter-comm}
  \end{subfigure}

    \begin{subfigure}{0.45\textwidth}
    \centering
    \showaxiom{facet-op}{2.5cm}{1.5cm}
    \caption{Disjoint filters and operations commute}
    \label{eq:filter-op}
  \end{subfigure}
  \begin{subfigure}{0.45\textwidth}
    \centering
    \showaxiom{facet-copy}{2.5cm}{1.5cm}
    \caption{Copying and filtering commute}
    \label{eq:filter-copy}
  \end{subfigure}

  \begin{subfigure}{0.95\textwidth}
    \centering
    \def\svgscale{0.8}
  \begin{tabular}{>{\centering\arraybackslash} m{3cm} E >{\centering\arraybackslash} m{3cm} E >{\centering\arraybackslash} m{3cm}}
\executeiffilenewer{diags/facet-discard-1.svg}{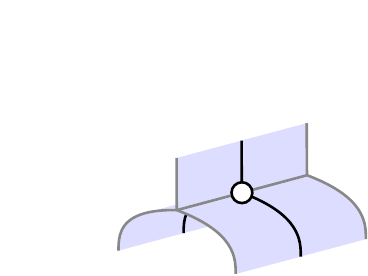}%
{inkscape -z -D --file=diags/facet-discard-1.svg%
--export-pdf=diags/facet-discard-1.pdf --export-latex}%
\begingroup%
  \makeatletter%
  \providecommand\color[2][]{%
    \errmessage{(Inkscape) Color is used for the text in Inkscape, but the package 'color.sty' is not loaded}%
    \renewcommand\color[2][]{}%
  }%
  \providecommand\transparent[1]{%
    \errmessage{(Inkscape) Transparency is used (non-zero) for the text in Inkscape, but the package 'transparent.sty' is not loaded}%
    \renewcommand\transparent[1]{}%
  }%
  \providecommand\rotatebox[2]{#2}%
  \newcommand*\fsize{\dimexpr\f@size pt\relax}%
  \newcommand*\lineheight[1]{\fontsize{\fsize}{#1\fsize}\selectfont}%
  \ifx\svgwidth\undefined%
    \setlength{\unitlength}{105.75000241bp}%
    \ifx\svgscale\undefined%
      \relax%
    \else%
      \setlength{\unitlength}{\unitlength * \real{\svgscale}}%
    \fi%
  \else%
    \setlength{\unitlength}{\svgwidth}%
  \fi%
  \global\let\svgwidth\undefined%
  \global\let\svgscale\undefined%
  \makeatother%
  \begin{picture}(1,0.74969828)%
    \lineheight{1}%
    \setlength\tabcolsep{0pt}%
    \put(0,0){\includegraphics[width=\unitlength,page=1]{facet-discard-1.pdf}}%
    \put(0.55248228,0.22495553){\makebox(0,0)[lt]{\lineheight{1.25}\smash{\begin{tabular}[t]{l}$f$\end{tabular}}}}%
    \put(0,0){\includegraphics[width=\unitlength,page=2]{facet-discard-1.pdf}}%
    \put(0.31241136,0.55665765){\makebox(0,0)[lt]{\lineheight{1.25}\smash{\begin{tabular}[t]{l}$f$\end{tabular}}}}%
    \put(0,0){\includegraphics[width=\unitlength,page=3]{facet-discard-1.pdf}}%
  \end{picture}%
\endgroup%
 &
    $=$ &
    \def\svgscale{0.8}
\executeiffilenewer{diags/facet-discard-2.svg}{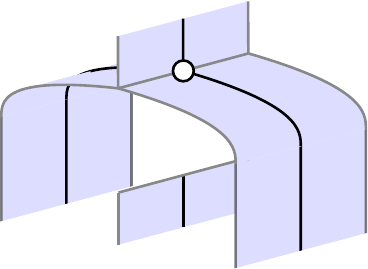}%
{inkscape -z -D --file=diags/facet-discard-2.svg%
--export-pdf=diags/facet-discard-2.pdf --export-latex}%
\begingroup%
  \makeatletter%
  \providecommand\color[2][]{%
    \errmessage{(Inkscape) Color is used for the text in Inkscape, but the package 'color.sty' is not loaded}%
    \renewcommand\color[2][]{}%
  }%
  \providecommand\transparent[1]{%
    \errmessage{(Inkscape) Transparency is used (non-zero) for the text in Inkscape, but the package 'transparent.sty' is not loaded}%
    \renewcommand\transparent[1]{}%
  }%
  \providecommand\rotatebox[2]{#2}%
  \newcommand*\fsize{\dimexpr\f@size pt\relax}%
  \newcommand*\lineheight[1]{\fontsize{\fsize}{#1\fsize}\selectfont}%
  \ifx\svgwidth\undefined%
    \setlength{\unitlength}{105.75000201bp}%
    \ifx\svgscale\undefined%
      \relax%
    \else%
      \setlength{\unitlength}{\unitlength * \real{\svgscale}}%
    \fi%
  \else%
    \setlength{\unitlength}{\svgwidth}%
  \fi%
  \global\let\svgwidth\undefined%
  \global\let\svgscale\undefined%
  \makeatother%
  \begin{picture}(1,0.73100307)%
    \lineheight{1}%
    \setlength\tabcolsep{0pt}%
    \put(0,0){\includegraphics[width=\unitlength,page=1]{facet-discard-2.pdf}}%
    \put(0.39290781,0.53796243){\makebox(0,0)[lt]{\lineheight{1.25}\smash{\begin{tabular}[t]{l}$f$\end{tabular}}}}%
    \put(0,0){\includegraphics[width=\unitlength,page=2]{facet-discard-2.pdf}}%
  \end{picture}%
\endgroup%
 &
        $=$ &
    \def\svgscale{0.8}
\executeiffilenewer{diags/facet-discard-3.svg}{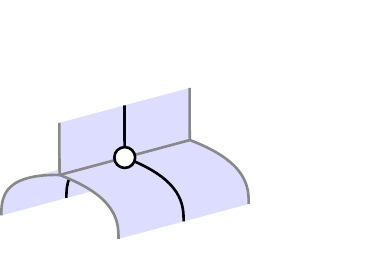}%
{inkscape -z -D --file=diags/facet-discard-3.svg%
--export-pdf=diags/facet-discard-3.pdf --export-latex}%
\begingroup%
  \makeatletter%
  \providecommand\color[2][]{%
    \errmessage{(Inkscape) Color is used for the text in Inkscape, but the package 'color.sty' is not loaded}%
    \renewcommand\color[2][]{}%
  }%
  \providecommand\transparent[1]{%
    \errmessage{(Inkscape) Transparency is used (non-zero) for the text in Inkscape, but the package 'transparent.sty' is not loaded}%
    \renewcommand\transparent[1]{}%
  }%
  \providecommand\rotatebox[2]{#2}%
  \newcommand*\fsize{\dimexpr\f@size pt\relax}%
  \newcommand*\lineheight[1]{\fontsize{\fsize}{#1\fsize}\selectfont}%
  \ifx\svgwidth\undefined%
    \setlength{\unitlength}{105.75000275bp}%
    \ifx\svgscale\undefined%
      \relax%
    \else%
      \setlength{\unitlength}{\unitlength * \real{\svgscale}}%
    \fi%
  \else%
    \setlength{\unitlength}{\svgwidth}%
  \fi%
  \global\let\svgwidth\undefined%
  \global\let\svgscale\undefined%
  \makeatother%
  \begin{picture}(1,0.71518746)%
    \lineheight{1}%
    \setlength\tabcolsep{0pt}%
    \put(0,0){\includegraphics[width=\unitlength,page=1]{facet-discard-3.pdf}}%
    \put(0.23333335,0.28618939){\makebox(0,0)[lt]{\lineheight{1.25}\smash{\begin{tabular}[t]{l}$f$\end{tabular}}}}%
    \put(0,0){\includegraphics[width=\unitlength,page=2]{facet-discard-3.pdf}}%
    \put(0.47198583,0.52214683){\makebox(0,0)[lt]{\lineheight{1.25}\smash{\begin{tabular}[t]{l}$f$\end{tabular}}}}%
    \put(0,0){\includegraphics[width=\unitlength,page=3]{facet-discard-3.pdf}}%
  \end{picture}%
\endgroup%

  \end{tabular}
    \caption{Filters do not modify data}
  \end{subfigure}
  \caption{Axioms of $\mathcal{F}$}
  \label{fig:axioms-f}
\end{figure}

\begin{defi}
  Let $\mathcal{F}$ be the free co-cartesian category generated as follows. We denote by $[A_1, \dots, A_n]$ the product of objects $A_1, \dots, A_n$ in $\mathcal{F}$ to distinguish it from
  the product $\times$ in $\mathcal{E}$. For each object $T \in \mathcal{E}$, $[T] \in \mathcal{F}$ is a generator. Morphism generators are:
  \begin{enumerate}[label=(\roman*)]
  \item For each morphism $\alpha \in \mathcal{E}(T,U)$, there is a generator $[\alpha]: [T] \rightarrow [U]$.
  \item For each filter $f$ and object $U \in \mathcal{E}$, there is a generator $[f \times U] : [T_f \times U] \rightarrow [T_f \times U, T_f \times U]$.
  \end{enumerate}
   For each object $T \in \mathcal{E}$, we call $J_T : [T, T] \rightarrow [T]$ and $E_T : [] \rightarrow [T]$ the comultiplication and counit provided by the co-cartesian structure.

  The axioms satisfied by these generators are stated graphically in Figure~\ref{fig:axioms-f}, with the notations introduced in Figure~\ref{fig:3d-notations}. In addition to these axioms, we require that $[g] \circ [f] = [g \circ f]$ (which is tautological graphically). In other words, $\mathcal{E}$ embeds into $\mathcal{F}$ functorially (but that functor is not monoidal).
\end{defi}

The definition above can be interpreted intuitively as follows. An
object in $\mathcal{E}$ represents the schema of a table (the list of
types of its columns). An object of $\mathcal{F}$ is a list of objects
of $\mathcal{E}$, so it represents a list of table schemata.  As will
be made clear by the semantics defined in the next section, a morphism
in $\mathcal{F} : [U, V] \mapsto [W, Z]$ should be thought of as a
function mapping disjoint tables of respective schemata $U$ and $V$ to
disjoint tables of respective schemata $W$ or $Z$, and row-wise so:
depending on its values, a row can end up in either of the output
tables. This makes it therefore possible to represent filters as
morphisms triaging rows to disjoint tables. A filter $[f \times U]$
operates on tables of schema $T_f \times U$, and only reads values
from the first component to determine whether to send the row to the
first or second output table. This treatment of a boolean predicate $A
\rightarrow 2$ as a morphism $A \rightarrow A + A$ is similar to that
of effectus theory~\cite{cho2015introduction}. The comultiplication
$J_T$ is a union, merges two tables of identical schemata
together.\footnote{In this model, row order does not matter in this
  model: tables are sets of rows.} The counit $E_T$ is the empty
table.

Given the two nested list structures in objects of $\mathcal{F}$, it
is natural to represent them as two-dimensional objects, and morphisms
of $\mathcal{F}$ become three-dimensional objects, as shown in
Figure~\ref{fig:3d-notations}. Figure~\ref{fig:axioms-f} states the
relations satisfied by these generators using this convention.

OpenRefine workflows with filters can be represented by morphisms $\mathcal{F}$.
For the converse, we first show that morphisms of $\mathcal{F}$ can
be represented in normal form thanks to the following decomposition.

\begin{lemmarep} \label{lemma:f-decomposition}
  Let $m \in \mathcal{F}([A], [B])$ be a morphism with one input sheet
  and one output sheet. There exists a decomposition
  $m = z \circ y \circ x \circ [w]$ such that $w \in \mathcal{E}$, $x$ only contains
  filters, $y$ only contains discarding morphisms, and $z$ only contains unions.
\end{lemmarep}
\begin{proof}
  First, any empty tables $E-T$ in the diagram can be eliminated as
  co-cartesian units, just like discarding morphisms can be eliminated
  in the cartesian case (Section~\ref{sec:first-model}).

  We then move all operations, copy morphisms and exchanges in $\mathcal{E}$ up to the first sheet.
  Operations and copy morphisms can be moved past unions and empty tables
  by the properties of the co-cartesian structure.
  Although Equation~\ref{eq:filter-op}
  can only be used for operations and filters applied to disjoint columns, it can be
  combined with Equation~\ref{eq:filter-copy} to commute any operation and filter,
  possibly leaving discarding morphisms behind:

  \noindent
  \begin{tabular}{H E G E G E G E F}
    \def\svgscale{0.7}
\executeiffilenewer{diags/proof-filter-op-1.svg}{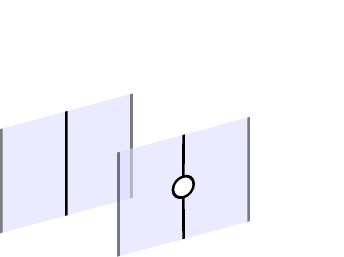}%
{inkscape -z -D --file=diags/proof-filter-op-1.svg%
--export-pdf=diags/proof-filter-op-1.pdf --export-latex}%
\begingroup%
  \makeatletter%
  \providecommand\color[2][]{%
    \errmessage{(Inkscape) Color is used for the text in Inkscape, but the package 'color.sty' is not loaded}%
    \renewcommand\color[2][]{}%
  }%
  \providecommand\transparent[1]{%
    \errmessage{(Inkscape) Transparency is used (non-zero) for the text in Inkscape, but the package 'transparent.sty' is not loaded}%
    \renewcommand\transparent[1]{}%
  }%
  \providecommand\rotatebox[2]{#2}%
  \newcommand*\fsize{\dimexpr\f@size pt\relax}%
  \newcommand*\lineheight[1]{\fontsize{\fsize}{#1\fsize}\selectfont}%
  \ifx\svgwidth\undefined%
    \setlength{\unitlength}{99.01142967bp}%
    \ifx\svgscale\undefined%
      \relax%
    \else%
      \setlength{\unitlength}{\unitlength * \real{\svgscale}}%
    \fi%
  \else%
    \setlength{\unitlength}{\svgwidth}%
  \fi%
  \global\let\svgwidth\undefined%
  \global\let\svgscale\undefined%
  \makeatother%
  \begin{picture}(1,0.74666709)%
    \lineheight{1}%
    \setlength\tabcolsep{0pt}%
    \put(0,0){\includegraphics[width=\unitlength,page=1]{proof-filter-op-1.pdf}}%
    \put(0.60902064,0.22334847){\rotatebox{15}{\makebox(0,0)[lt]{\lineheight{1.25}\smash{\begin{tabular}[t]{l}$\alpha$\end{tabular}}}}}%
    \put(0,0){\includegraphics[width=\unitlength,page=2]{proof-filter-op-1.pdf}}%
    \put(0.24921368,0.54048839){\makebox(0,0)[lt]{\lineheight{1.25}\smash{\begin{tabular}[t]{l}$f$\end{tabular}}}}%
    \put(0,0){\includegraphics[width=\unitlength,page=3]{proof-filter-op-1.pdf}}%
  \end{picture}%
\endgroup%
 &
    $=$ &
    \def\svgscale{0.7}
\executeiffilenewer{diags/proof-filter-op-2.svg}{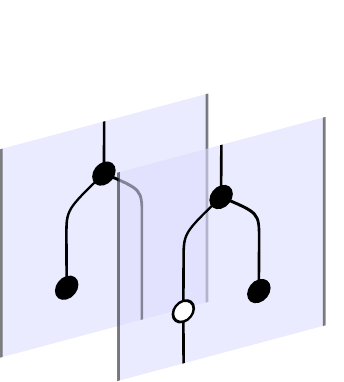}%
{inkscape -z -D --file=diags/proof-filter-op-2.svg%
--export-pdf=diags/proof-filter-op-2.pdf --export-latex}%
\begingroup%
  \makeatletter%
  \providecommand\color[2][]{%
    \errmessage{(Inkscape) Color is used for the text in Inkscape, but the package 'color.sty' is not loaded}%
    \renewcommand\color[2][]{}%
  }%
  \providecommand\transparent[1]{%
    \errmessage{(Inkscape) Transparency is used (non-zero) for the text in Inkscape, but the package 'transparent.sty' is not loaded}%
    \renewcommand\transparent[1]{}%
  }%
  \providecommand\rotatebox[2]{#2}%
  \newcommand*\fsize{\dimexpr\f@size pt\relax}%
  \newcommand*\lineheight[1]{\fontsize{\fsize}{#1\fsize}\selectfont}%
  \ifx\svgwidth\undefined%
    \setlength{\unitlength}{99.01142967bp}%
    \ifx\svgscale\undefined%
      \relax%
    \else%
      \setlength{\unitlength}{\unitlength * \real{\svgscale}}%
    \fi%
  \else%
    \setlength{\unitlength}{\svgwidth}%
  \fi%
  \global\let\svgwidth\undefined%
  \global\let\svgscale\undefined%
  \makeatother%
  \begin{picture}(1,1.10852324)%
    \lineheight{1}%
    \setlength\tabcolsep{0pt}%
    \put(0,0){\includegraphics[width=\unitlength,page=1]{proof-filter-op-2.pdf}}%
    \put(0.60902064,0.22334847){\rotatebox{15}{\makebox(0,0)[lt]{\lineheight{1.25}\smash{\begin{tabular}[t]{l}$\alpha$\end{tabular}}}}}%
    \put(0,0){\includegraphics[width=\unitlength,page=2]{proof-filter-op-2.pdf}}%
    \put(0.35904948,0.87291413){\makebox(0,0)[lt]{\lineheight{1.25}\smash{\begin{tabular}[t]{l}$f$\end{tabular}}}}%
    \put(0,0){\includegraphics[width=\unitlength,page=3]{proof-filter-op-2.pdf}}%
  \end{picture}%
\endgroup%
 &
    $=$ &
    \def\svgscale{0.7}
\executeiffilenewer{diags/proof-filter-op-3.svg}{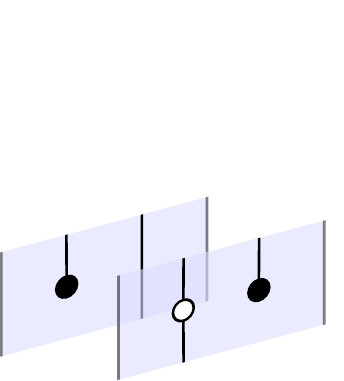}%
{inkscape -z -D --file=diags/proof-filter-op-3.svg%
--export-pdf=diags/proof-filter-op-3.pdf --export-latex}%
\begingroup%
  \makeatletter%
  \providecommand\color[2][]{%
    \errmessage{(Inkscape) Color is used for the text in Inkscape, but the package 'color.sty' is not loaded}%
    \renewcommand\color[2][]{}%
  }%
  \providecommand\transparent[1]{%
    \errmessage{(Inkscape) Transparency is used (non-zero) for the text in Inkscape, but the package 'transparent.sty' is not loaded}%
    \renewcommand\transparent[1]{}%
  }%
  \providecommand\rotatebox[2]{#2}%
  \newcommand*\fsize{\dimexpr\f@size pt\relax}%
  \newcommand*\lineheight[1]{\fontsize{\fsize}{#1\fsize}\selectfont}%
  \ifx\svgwidth\undefined%
    \setlength{\unitlength}{99.01142967bp}%
    \ifx\svgscale\undefined%
      \relax%
    \else%
      \setlength{\unitlength}{\unitlength * \real{\svgscale}}%
    \fi%
  \else%
    \setlength{\unitlength}{\svgwidth}%
  \fi%
  \global\let\svgwidth\undefined%
  \global\let\svgscale\undefined%
  \makeatother%
  \begin{picture}(1,1.10575064)%
    \lineheight{1}%
    \setlength\tabcolsep{0pt}%
    \put(0,0){\includegraphics[width=\unitlength,page=1]{proof-filter-op-3.pdf}}%
    \put(0.60902064,0.22334847){\rotatebox{15}{\makebox(0,0)[lt]{\lineheight{1.25}\smash{\begin{tabular}[t]{l}$\alpha$\end{tabular}}}}}%
    \put(0,0){\includegraphics[width=\unitlength,page=2]{proof-filter-op-3.pdf}}%
    \put(0.46888529,0.59934922){\makebox(0,0)[lt]{\lineheight{1.25}\smash{\begin{tabular}[t]{l}$f$\end{tabular}}}}%
    \put(0,0){\includegraphics[width=\unitlength,page=3]{proof-filter-op-3.pdf}}%
  \end{picture}%
\endgroup%
 &
        $=$ &
    \def\svgscale{0.7}
\executeiffilenewer{diags/proof-filter-op-4.svg}{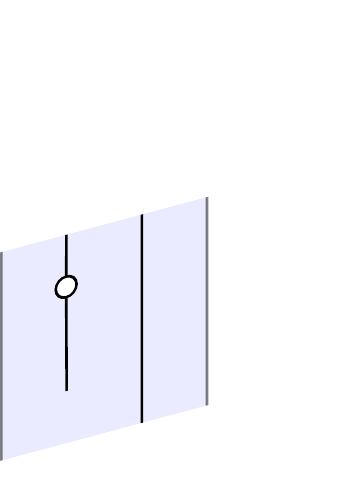}%
{inkscape -z -D --file=diags/proof-filter-op-4.svg%
--export-pdf=diags/proof-filter-op-4.pdf --export-latex}%
\begingroup%
  \makeatletter%
  \providecommand\color[2][]{%
    \errmessage{(Inkscape) Color is used for the text in Inkscape, but the package 'color.sty' is not loaded}%
    \renewcommand\color[2][]{}%
  }%
  \providecommand\transparent[1]{%
    \errmessage{(Inkscape) Transparency is used (non-zero) for the text in Inkscape, but the package 'transparent.sty' is not loaded}%
    \renewcommand\transparent[1]{}%
  }%
  \providecommand\rotatebox[2]{#2}%
  \newcommand*\fsize{\dimexpr\f@size pt\relax}%
  \newcommand*\lineheight[1]{\fontsize{\fsize}{#1\fsize}\selectfont}%
  \ifx\svgwidth\undefined%
    \setlength{\unitlength}{99.01142967bp}%
    \ifx\svgscale\undefined%
      \relax%
    \else%
      \setlength{\unitlength}{\unitlength * \real{\svgscale}}%
    \fi%
  \else%
    \setlength{\unitlength}{\svgwidth}%
  \fi%
  \global\let\svgwidth\undefined%
  \global\let\svgscale\undefined%
  \makeatother%
  \begin{picture}(1,1.40874596)%
    \lineheight{1}%
    \setlength\tabcolsep{0pt}%
    \put(0,0){\includegraphics[width=\unitlength,page=1]{proof-filter-op-4.pdf}}%
    \put(0.2681509,0.59451774){\rotatebox{15}{\makebox(0,0)[lt]{\lineheight{1.25}\smash{\begin{tabular}[t]{l}$\alpha$\end{tabular}}}}}%
    \put(0,0){\includegraphics[width=\unitlength,page=2]{proof-filter-op-4.pdf}}%
    \put(0.60902064,0.52634379){\rotatebox{15}{\makebox(0,0)[lt]{\lineheight{1.25}\smash{\begin{tabular}[t]{l}$\alpha$\end{tabular}}}}}%
    \put(0,0){\includegraphics[width=\unitlength,page=3]{proof-filter-op-4.pdf}}%
    \put(0.46888529,0.90234454){\makebox(0,0)[lt]{\lineheight{1.25}\smash{\begin{tabular}[t]{l}$f$\end{tabular}}}}%
    \put(0,0){\includegraphics[width=\unitlength,page=4]{proof-filter-op-4.pdf}}%
  \end{picture}%
\endgroup%
 &
        $=$ &
    \def\svgscale{0.7}
\executeiffilenewer{diags/proof-filter-op-5.svg}{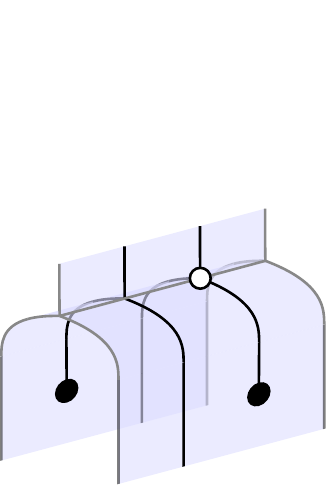}%
{inkscape -z -D --file=diags/proof-filter-op-5.svg%
--export-pdf=diags/proof-filter-op-5.pdf --export-latex}%
\begingroup%
  \makeatletter%
  \providecommand\color[2][]{%
    \errmessage{(Inkscape) Color is used for the text in Inkscape, but the package 'color.sty' is not loaded}%
    \renewcommand\color[2][]{}%
  }%
  \providecommand\transparent[1]{%
    \errmessage{(Inkscape) Transparency is used (non-zero) for the text in Inkscape, but the package 'transparent.sty' is not loaded}%
    \renewcommand\transparent[1]{}%
  }%
  \providecommand\rotatebox[2]{#2}%
  \newcommand*\fsize{\dimexpr\f@size pt\relax}%
  \newcommand*\lineheight[1]{\fontsize{\fsize}{#1\fsize}\selectfont}%
  \ifx\svgwidth\undefined%
    \setlength{\unitlength}{93.75000275bp}%
    \ifx\svgscale\undefined%
      \relax%
    \else%
      \setlength{\unitlength}{\unitlength * \real{\svgscale}}%
    \fi%
  \else%
    \setlength{\unitlength}{\svgwidth}%
  \fi%
  \global\let\svgwidth\undefined%
  \global\let\svgscale\undefined%
  \makeatother%
  \begin{picture}(1,1.48780744)%
    \lineheight{1}%
    \setlength\tabcolsep{0pt}%
    \put(0,0){\includegraphics[width=\unitlength,page=1]{proof-filter-op-5.pdf}}%
    \put(0.49520001,0.63298583){\makebox(0,0)[lt]{\lineheight{1.25}\smash{\begin{tabular}[t]{l}$f$\end{tabular}}}}%
    \put(0,0){\includegraphics[width=\unitlength,page=2]{proof-filter-op-5.pdf}}%
    \put(0.46160003,0.91220319){\rotatebox{15}{\makebox(0,0)[lt]{\lineheight{1.25}\smash{\begin{tabular}[t]{l}$\alpha$\end{tabular}}}}}%
  \end{picture}%
\endgroup%

  \end{tabular}

  \noindent This lets us push all operations up, obtaining the first part of the factorization: $m = \phi \circ [w]$ with $w \in \mathcal{E}$ and $\phi$ consists of filters, unions, discarding morphisms and exchanges in $\mathcal{F}$.
 
  Unions can be moved
  down by naturality, obtaining $m = z \circ \phi' \circ [w]$ where $\phi'$ only consists of filters, discarding morphisms and exchanges in $\mathcal{F}$.
  Then, all exchanges in $\phi'$ can be moved down by naturality and absorbed by $z$. Finally, all discarding morphisms can be moved down past the filters using Equation~\ref{eq:filter-op}.
\end{proof}

\noindent This decomposition can be used to show that all such
morphisms arise from OpenRefine workflows, despite the fact that some
generators cannot be interpreted as such individually. As stated, this
lemma does not provide normal forms yet, as the order of filters in
$x$ is not determined.  We will see in the proof of
Theorem~\ref{thm:completeness} how this can be addressed.

\section{Semantics and completeness}

We can give set-valued semantics to $\mathcal{E}$ and $\mathcal{F}$ and obtain
completeness theorems for our axiomatization of OpenRefine workflows.

\begin{defi}
  A \emph{valuation} $V$ is given by:
  \begin{enumerate}[label=(\roman*)]
  \item a set $V(T)$ for each basic datatype $T \in \mathcal{E}$;
  \item a function $V(\alpha) : V(A) \rightarrow V(B)$ for each generator $\alpha \in \mathcal{E}(A,B)$, where $V(A)$ is the cartesian product of the valuations of the basic types in $A$;
  \item a subset $V(f) \subset V(T_f)$ for each filter $f$.
  \end{enumerate}
\end{defi}

\noindent Any valuation $V$ defines a functor $V^* : \mathcal{F} \rightarrow \mathbf{Set}$ as follows:
\begin{align*}
  V^*([A \times \dots \times B, \dots, C \times \dots \times D]) &= (V(A) \times \dots \times V(B) \sqcup \dots \sqcup (V(C) \times \dots \times V(D)) \\
  V^*([\alpha]) &= V(\alpha) \\
  V^*([f \times U]) &= ((x,u) \mapsto \inj_i(x,u)) \text{ with $i = 1$ if $x \in V(f)$ else $i=2$} \\
  V^*([J_T]) &= (\inj_i(x) \mapsto x) \\
  V^*([E_T]) &= (the initial morphism from the empty set)
\end{align*}
Using the decomposition of Lemma~\ref{lemma:f-decomposition}, we can then show the completeness of our axiomatization for these semantics:
\begin{thmrep} \label{thm:completeness}
  Let $d, d' \in \mathcal{F}([A],[B])$ be diagrams. Then $d = d'$ by the axioms of $\mathcal{F}$ if and only if $V^*(d) = V^*(d')$ for any valuation $V$.
\end{thmrep}
The proof of this theorem is given in appendix. Broadly speaking, it goes by building a valuation where values are syntactic terms, such that a value encodes its entire own history through the processing pipeline. These syntactic values are associated with contexts which record the validity of filter expressions. The decomposition of Lemma~\ref{lemma:f-decomposition} is then used to compute normal forms for diagrams, which can be related to the evaluation of the diagram with the syntactic valuation.
These normal forms can be computed using a simple diagramatic rewriting strategy, so this also
solves the word problem for this signature.

\begin{proof}
  We can check that all equations of Figure~\ref{fig:axioms-f} preserve the semantics under any valuation, so if two diagrams are equivalent up to these axioms, then their interpretations are equal.
  For the converse, let us first introduce a few notions. We use a countable set of variables $V = \{x_1, x_2, x_3, \dots \}$.
  
  \begin{defi}
    The set $\Theta$ of \textbf{terms} is defined inductively: it
    contains the variables $V$, and for each an operation symbol $\alpha
    \in O$ of input arity $n$ and output arity $p$, it contains the
    terms from $\alpha(t_1, \dots, t_n)[1]$ to $\alpha(t_1, \dots, t_n)[p]$. These
    terms represent the projections of the operation applied to the
    input terms.
  \end{defi}
  The set $\Theta_n$ of terms over $n$ variables is the set of terms
  where only variables from $\{ x_1, \dots, x_n \}$ are used.  Given
  $t \in \Theta_n$ and $u_1, \dots, u_n \in \Theta_m$ we can
  substitute simultaneously all the $x_i$ by $u_i$, which we denote by
  $t[u_1, \dots, u_n]$.  For instance, let $t = \alpha(\beta(x_1,
  x_3)[2], x_1)[1]$ and $u_1 = x_3$, $u_2 = x_4$ and $u_3 = \gamma(x_1)[3]$.
  Then $t[u_1, u_2, u_3] = \alpha(\beta(x_3, \gamma(x_1)[3])[2], x_3)[1]$.
 
  \begin{defi}
    An \textbf{atomic filter formula} (\textbf{AFF}) over $n$ variables is
    given by a filter symbol $f$ and terms $t_1, \dots, t_a \in \Theta_n$ where $a$
    is the arity of $f$. It is denoted by $f(t_1, \dots, t_a)$ and
    represents the boolean condition evaluated on the given terms.
  \end{defi}
  We denote by $\Phi$ the set of all atomic filter formulae.  Similarly, $\Phi_n$
  is the set of AFF over $n$ variables.
  \begin{defi}
    A \textbf{conjunctive filter formula} (\textbf{CFF}) over $n$ variables is a given by a finite set
    $A \subset \Phi_n \times \mathbb{B}$ of pairs of atomic filter
    formulae and booleans, called clauses, such that no atomic filter
    formula appears with both booleans. Such a set represents the
    conjunction of all its clauses, negated when their associated
    boolean is false.

    Two CFF $A$ and $B$ are disjoint if they contain
    the same atomic filter formula with opposite booleans.
    Otherwise, we can form the conjuction $A \land B$, which is the CFF with clauses $A \cup B$.
  \end{defi}
  We denote by $\Delta$ the set of CFF and $\Delta_n$ that of those over $n$ variables.
  A CFF is represented as a conjuctive clause in boolean logic, such
  as $f(x_1, \alpha(x_3, x_2)[1]) \land \bar{g}(x_3)$.
  \begin{defi}
    A \textbf{truth table} $t$ on $n$ variables and $p$ outputs, denoted by $t: n \rightarrow p$, is a
    finite set of cases $c \in \Delta_n \times \Theta_n^p$, such that all the CFF are pairwise disjoint. This represents possible values for an object, depending on the evaluation of some filters.
    
    Truth tables $t, t'$ both on $n$ variables and $p$ outputs are
    \textbf{disjoint} if all the CFF involved are pairwise disjoint.
    The \textbf{union} of two disjoint truth tables $t,t'$, denoted by
    $t \cup t'$, is the union of their cases.  The
    \textbf{composition} of truth tables $t : n \rightarrow p$ and
    $t': p \rightarrow q$, denoted by $t ; t'$, is formed of the cases
    $(c_i \land c'_j, u'_{j,1}[u_{i,1}, \dots, u_{i,p}], \dots,
    u'_{j,q}[u_{i,1}, \dots, u_{i,p}])$ for all $(c_i, u_i) \in t$ and
    $(c'_j,u'_j)$ such that $c_i$ and $c_j$ are compatible.

    A collection of truth tables $(t_i : n \rightarrow p)_i$ forms a \textbf{partition} if the
    CFF in them are all disjoint and their disjunction is a tautology.
    The \textbf{projection} of a truth table $t$ with $p$ outputs on its $k$-th component, $1 \leq k \leq k$, denoted by $t[k]$, is given by the cases $(c_i, u_{i,k})$ for $(c_i, u_i) \in t$.
    The \textbf{product} of truth tables $t : n \rightarrow p$ and $t' : n \rightarrow q$, denoted by $t \otimes t' : n \rightarrow p + q$, is given by the cases $(c_i \land c'_j, u_i, u'_j)$ for all $(c_i, u_i) \in t$ and $(c'_j, u'_j) \in t'$ such that $c_i$ and $c_j$ are compatible.
    Two truth tables $t, t' : n \rightarrow p$ are \textbf{equivalent}, denoted by $t \sim t'$,
    if all the cases in $t \otimes t'$ have value tuples of the form $(v_1, \dots, v_p, v_1, \dots, v_p)$.
  \end{defi}
  \noindent One can check that all the properties and operations on truth tables defined
  above respect the equivalence relation $\sim$: we will therefore work up to this equivalence
  in the sequel. We can represent truth tables by their list of cases:
  \begin{align*}
    f(x_1) \land g(\gamma(x_2)[1]) &\mapsto (x_3, \alpha(x_1, x_1)[1]) \\
    f(x_1) \land \bar{g}(\gamma(x_2)[1]) & \mapsto (x_2, \alpha(x_1, x_1)[1]) \\
    \bar{f}(x_1) &\mapsto (\beta(x_2, x_3)[2], x_1)
  \end{align*}
  With the syntactic objects just defined, we can now define semantics
  for $\mathcal{F}$ that are independent from any valuation. The
  morphisms will be families of truth tables, which can interpret the
  generators of $\mathcal{F}$.
  
  \begin{defi}
    The category $\mathcal{T}$ is a symmetric monoidal category with $\Ob(\mathcal{T}) = \mathbb{N}^*$ (lists of natural numbers) and
    where the monoidal product is given by concatenation.

    A morphism $t \in \mathcal{T}([a_1, \dots, a_n], [b_1, \dots, b_m])$ is a collection $t_{i,j}$ of truth tables, $1 \leq i \leq n$ and $1 \leq j \leq m$, such that $t_{i,j}$ is of type $a_i \rightarrow b_j$, and for each $i$, $(t_{i,j})_j$ is a partition. Furthermore we require that $m > 0$ unless $n=0$.

    Given morphisms $t : [a_1, \dots, a_n] \rightarrow [b_1, \dots,
      b_m]$ and $u : [b_1, \dots, b_m] \rightarrow [c_1, \dots, c_p]$, the composite $t; u$
    is given by $(t ; u)_{i,k} = \bigcup_{1 \leq j \leq m} (t_{i,j} ; u_{j,k})$.

    The tensor product of $t : [a_1, \dots, a_n] \rightarrow [b_1, \dots, b_m]$ and $u : [c_1, \dots, c_p] \rightarrow [d_1, \dots, d_q]$ is the morphism $v : [a_1, \dots, a_n, c_1, \dots, c_p] \rightarrow [b_1, \dots, b_m, d_1, \dots, d_q]$ defined by $v_{i,j} = t_{i,j}$ for $i \leq n$ and $j \leq p$, $v_{i,j}= u_{i-n,j-m}$ for $i > n$ and $j > p$, and the empty truth table otherwise.

    The identity $1$ on $[a_1 \dots, a_n]$ is given by $1_i = \top \mapsto (x_1, \dots, x_n)$.
  \end{defi}

\begin{figure}[h]
  \centering
  \begin{subfigure}{0.59\textwidth}
    \centering
    \begin{tikzpicture}
      \node at (0,0) {%
\executeiffilenewer{diags/operation-def.svg}{diags/operation-def.pdf}%
{inkscape -z -D --file=diags/operation-def.svg%
--export-pdf=diags/operation-def.pdf --export-latex}%
};
      \node at (1.5,0) {$\mapsto$};
      \draw[-latex] (5,1) -- (5,0);
      \node at (5,-.5) {$(\top \mapsto (\alpha(x_1, x_2)[1], \alpha(x_1, x_2)[2]))$};
    \end{tikzpicture}
  \end{subfigure}
  \begin{subfigure}{0.75\textwidth}
    \centering
    \begin{tikzpicture}
      \node at (0,0) {%
\executeiffilenewer{diags/facet-def.svg}{diags/facet-def.pdf}%
{inkscape -z -D --file=diags/facet-def.svg%
--export-pdf=diags/facet-def.pdf --export-latex}%
\input{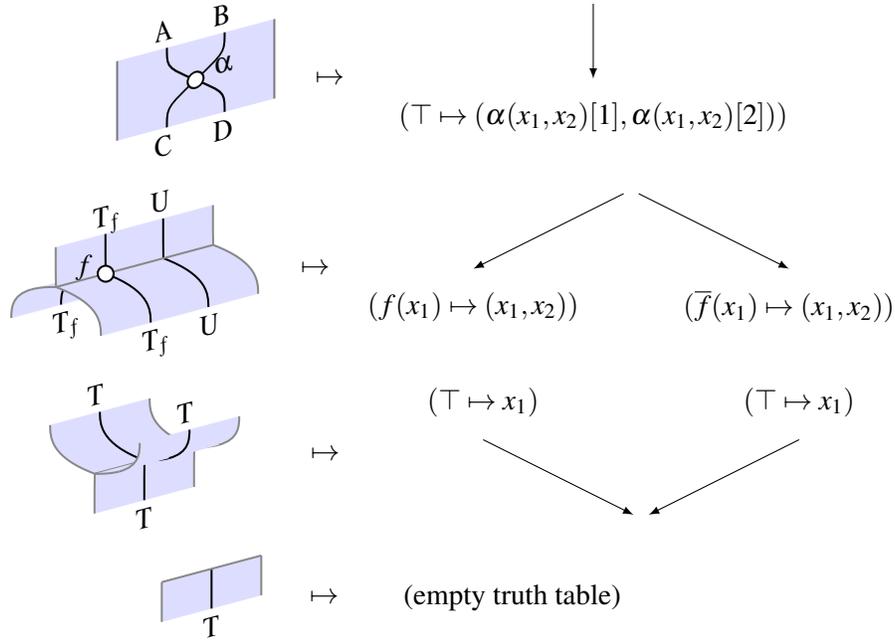}%
};

      \draw[-latex] (6.5,1) -- (4.5,0);
      \draw[-latex] (6.7,1) -- (8.7,0);
      \node at (2.4,0) {$\mapsto$};
      \node at (4.5,-.5) {$(f(x_1) \mapsto (x_1,x_2))$};
      \node at (8.7,-.5) {$(\bar{f}(x_1) \mapsto (x_1,x_2))$};

    \end{tikzpicture}
  \end{subfigure}
  \begin{subfigure}{0.7\textwidth}
    \centering
    \begin{tikzpicture}
      \node at (0,0) {%
\executeiffilenewer{diags/join-def.svg}{diags/join-def.pdf}%
{inkscape -z -D --file=diags/join-def.svg%
--export-pdf=diags/join-def.pdf --export-latex}%
};
      \node at (2.4,0) {$\mapsto$};
            \begin{scope}[yshift=-.8cm]
      \draw[-latex] (4.5,1) -- (6.5, 0);
      \draw[-latex] (8.7,1) -- (6.7, 0);

      \node at (4.5,1.5) {$(\top \mapsto x_1)$};
      \node at (8.7,1.5) {$(\top \mapsto x_1)$};
            \end{scope}
    \end{tikzpicture}
  \end{subfigure}
  \begin{subfigure}{0.5\textwidth}
    \begin{tikzpicture}
      \node at (0,0) {%
\executeiffilenewer{diags/empty-def.svg}{diags/empty-def.pdf}%
{inkscape -z -D --file=diags/empty-def.svg%
--export-pdf=diags/empty-def.pdf --export-latex}%
};
      \node at (1.5,0) {$\mapsto$};
      \node at (4,0) {(empty truth table)};
    \end{tikzpicture}
  \end{subfigure}
  \caption{Definition of $P : \mathcal{F} \rightarrow \mathcal{T}$}
  \label{fig:def-functor-P}
\end{figure}

\noindent There is a functor $P : \mathcal{F} \rightarrow \mathcal{T}$ defined on objects by $P([A_1 \times \dots \times A_n]) = [n]$ and on morphisms by Figure~\ref{fig:def-functor-P}. One can check that it respects the axioms of $\mathcal{F}$.

  \begin{lemma} \label{lemma:p-faithful}
    The functor $P$ is faithful.
  \end{lemma}
  
  \begin{nestedproof}
    We show this by relating the image $P(d)$ of a diagram to its decomposition
    given by Lemma~\ref{lemma:f-decomposition}. As such, this decomposition
    does not give a normal form, as the order of the filters remains unspecified.
    However, successive filters can be swapped freely:
\begin{figure}[H]
\vspace{-.2cm} 
\centering
  \begin{tabular}{G E I E I}
    \def\svgscale{0.7}
\executeiffilenewer{diags/proof-filter-comm-1.svg}{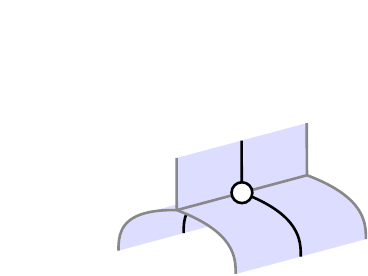}%
{inkscape -z -D --file=diags/proof-filter-comm-1.svg%
--export-pdf=diags/proof-filter-comm-1.pdf --export-latex}%
\begingroup%
  \makeatletter%
  \providecommand\color[2][]{%
    \errmessage{(Inkscape) Color is used for the text in Inkscape, but the package 'color.sty' is not loaded}%
    \renewcommand\color[2][]{}%
  }%
  \providecommand\transparent[1]{%
    \errmessage{(Inkscape) Transparency is used (non-zero) for the text in Inkscape, but the package 'transparent.sty' is not loaded}%
    \renewcommand\transparent[1]{}%
  }%
  \providecommand\rotatebox[2]{#2}%
  \newcommand*\fsize{\dimexpr\f@size pt\relax}%
  \newcommand*\lineheight[1]{\fontsize{\fsize}{#1\fsize}\selectfont}%
  \ifx\svgwidth\undefined%
    \setlength{\unitlength}{105.75000241bp}%
    \ifx\svgscale\undefined%
      \relax%
    \else%
      \setlength{\unitlength}{\unitlength * \real{\svgscale}}%
    \fi%
  \else%
    \setlength{\unitlength}{\svgwidth}%
  \fi%
  \global\let\svgwidth\undefined%
  \global\let\svgscale\undefined%
  \makeatother%
  \begin{picture}(1,0.74969828)%
    \lineheight{1}%
    \setlength\tabcolsep{0pt}%
    \put(0,0){\includegraphics[width=\unitlength,page=1]{proof-filter-comm-1.pdf}}%
    \put(0.55248228,0.22495553){\makebox(0,0)[lt]{\lineheight{1.25}\smash{\begin{tabular}[t]{l}$g$\end{tabular}}}}%
    \put(0,0){\includegraphics[width=\unitlength,page=2]{proof-filter-comm-1.pdf}}%
    \put(0.31241136,0.55665765){\makebox(0,0)[lt]{\lineheight{1.25}\smash{\begin{tabular}[t]{l}$f$\end{tabular}}}}%
    \put(0,0){\includegraphics[width=\unitlength,page=3]{proof-filter-comm-1.pdf}}%
  \end{picture}%
\endgroup%
 &
    $=$ &
    \def\svgscale{0.7}
\executeiffilenewer{diags/proof-filter-comm-2.svg}{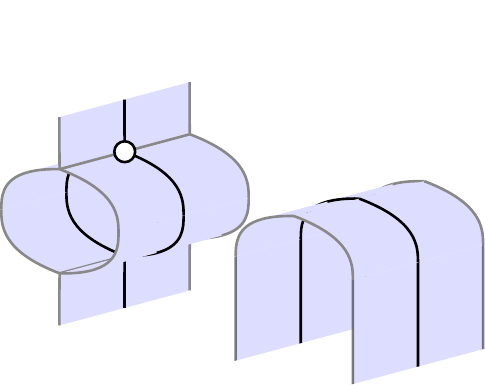}%
{inkscape -z -D --file=diags/proof-filter-comm-2.svg%
--export-pdf=diags/proof-filter-comm-2.pdf --export-latex}%
\begingroup%
  \makeatletter%
  \providecommand\color[2][]{%
    \errmessage{(Inkscape) Color is used for the text in Inkscape, but the package 'color.sty' is not loaded}%
    \renewcommand\color[2][]{}%
  }%
  \providecommand\transparent[1]{%
    \errmessage{(Inkscape) Transparency is used (non-zero) for the text in Inkscape, but the package 'transparent.sty' is not loaded}%
    \renewcommand\transparent[1]{}%
  }%
  \providecommand\rotatebox[2]{#2}%
  \newcommand*\fsize{\dimexpr\f@size pt\relax}%
  \newcommand*\lineheight[1]{\fontsize{\fsize}{#1\fsize}\selectfont}%
  \ifx\svgwidth\undefined%
    \setlength{\unitlength}{139.50000275bp}%
    \ifx\svgscale\undefined%
      \relax%
    \else%
      \setlength{\unitlength}{\unitlength * \real{\svgscale}}%
    \fi%
  \else%
    \setlength{\unitlength}{\svgwidth}%
  \fi%
  \global\let\svgwidth\undefined%
  \global\let\svgscale\undefined%
  \makeatother%
  \begin{picture}(1,0.79360985)%
    \lineheight{1}%
    \setlength\tabcolsep{0pt}%
    \put(0,0){\includegraphics[width=\unitlength,page=1]{proof-filter-comm-2.pdf}}%
    \put(0.17688174,0.48039088){\makebox(0,0)[lt]{\lineheight{1.25}\smash{\begin{tabular}[t]{l}$g$\end{tabular}}}}%
    \put(0,0){\includegraphics[width=\unitlength,page=2]{proof-filter-comm-2.pdf}}%
    \put(0.66075269,0.38361668){\makebox(0,0)[lt]{\lineheight{1.25}\smash{\begin{tabular}[t]{l}$g$\end{tabular}}}}%
    \put(0,0){\includegraphics[width=\unitlength,page=3]{proof-filter-comm-2.pdf}}%
    \put(0.41774195,0.64727259){\makebox(0,0)[lt]{\lineheight{1.25}\smash{\begin{tabular}[t]{l}$f$\end{tabular}}}}%
    \put(0,0){\includegraphics[width=\unitlength,page=4]{proof-filter-comm-2.pdf}}%
  \end{picture}%
\endgroup%
 &
    $=$ &
    \def\svgscale{0.7}
\executeiffilenewer{diags/proof-filter-comm-3.svg}{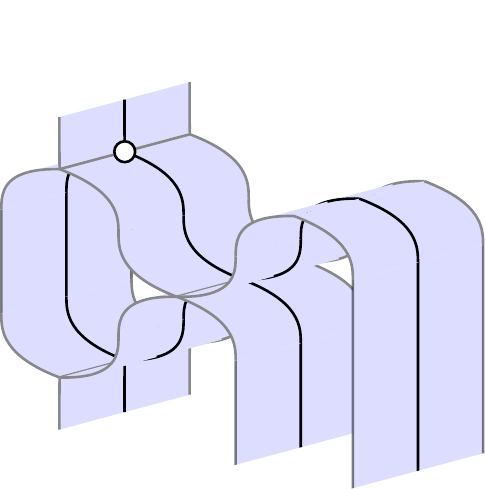}%
{inkscape -z -D --file=diags/proof-filter-comm-3.svg%
--export-pdf=diags/proof-filter-comm-3.pdf --export-latex}%
\begingroup%
  \makeatletter%
  \providecommand\color[2][]{%
    \errmessage{(Inkscape) Color is used for the text in Inkscape, but the package 'color.sty' is not loaded}%
    \renewcommand\color[2][]{}%
  }%
  \providecommand\transparent[1]{%
    \errmessage{(Inkscape) Transparency is used (non-zero) for the text in Inkscape, but the package 'transparent.sty' is not loaded}%
    \renewcommand\transparent[1]{}%
  }%
  \providecommand\rotatebox[2]{#2}%
  \newcommand*\fsize{\dimexpr\f@size pt\relax}%
  \newcommand*\lineheight[1]{\fontsize{\fsize}{#1\fsize}\selectfont}%
  \ifx\svgwidth\undefined%
    \setlength{\unitlength}{139.50000275bp}%
    \ifx\svgscale\undefined%
      \relax%
    \else%
      \setlength{\unitlength}{\unitlength * \real{\svgscale}}%
    \fi%
  \else%
    \setlength{\unitlength}{\svgwidth}%
  \fi%
  \global\let\svgwidth\undefined%
  \global\let\svgscale\undefined%
  \makeatother%
  \begin{picture}(1,1.0086636)%
    \lineheight{1}%
    \setlength\tabcolsep{0pt}%
    \put(0,0){\includegraphics[width=\unitlength,page=1]{proof-filter-comm-3.pdf}}%
    \put(0.17688174,0.69544463){\makebox(0,0)[lt]{\lineheight{1.25}\smash{\begin{tabular}[t]{l}$f$\end{tabular}}}}%
    \put(0,0){\includegraphics[width=\unitlength,page=2]{proof-filter-comm-3.pdf}}%
    \put(0.66075269,0.59867044){\makebox(0,0)[lt]{\lineheight{1.25}\smash{\begin{tabular}[t]{l}$f$\end{tabular}}}}%
    \put(0,0){\includegraphics[width=\unitlength,page=3]{proof-filter-comm-3.pdf}}%
    \put(0.41774195,0.86232635){\makebox(0,0)[lt]{\lineheight{1.25}\smash{\begin{tabular}[t]{l}$g$\end{tabular}}}}%
    \put(0,0){\includegraphics[width=\unitlength,page=4]{proof-filter-comm-3.pdf}}%
  \end{picture}%
\endgroup%

  \end{tabular}
\vspace{-.5cm}
\end{figure}
  
  \noindent Let us pick an arbitrary order on $\Phi$, the set of atomic filter
  formulae.  In a diagram $d$ decomposed by
  Lemma~\ref{lemma:f-decomposition} as $z \circ y \circ x \circ [w]$,
  Each occurence of a filter in $x$ can be associated with an AFF
  defined by the filter symbol for the filter and the term obtained
  from the wires read from $w$. Commuting filters as above does not
  change their corresponding AFF.  Therefore this determines an order
  on the filters occurring in $x$.  We can rearrange the filters so
  that $f$ appears above $g$ if their corresponding AFFs are ordered
  accordingly. This will add new exchanges and unions in $x$, but we
  can use Lemma~\ref{lemma:f-decomposition} a second time to push
  these to their part of the decomposition, as this procedure does not
  reorder the filters.

  The rest of the decomposition can be normalized too: unions can be normalized
  by associativity, and any discarding
  morphism that is present in all sheets of $y$ and discards a wire
  not read by any filter in $x$ can be pushed up into $w$, which
  can be normalized as a morphism in $\mathcal{E}$.

  From such a normalized decomposition, we can read out the truth table $P(d)$
  directly. Each sheet in $y$ corresponds to a case of $P(d)$, whose condition
  is determined by the conjunction of all the AFFs of the filters leading to it,
  with the appropriate boolean depending on the side of the filter they are on.
  Therefore, if $P(d) = P(d')$, then $d = d'$.
  \end{nestedproof}
  
  \begin{defi}
    The \textbf{syntactic valuation} $S$ is defined as follows. For each basic datype $T$, $S(T) = \Theta \times 2^\Phi + \{ \bot \}$.
    In other words, a value can be either a term together with a context of true atomic filter formulae, or an inconsistent value $\bot$.

\noindent For each facet $f$, $S(f) : (C,t) \mapsto f(t) \in C$: a facet is true if it belongs to the context.  
    
\noindent For each operation $\alpha : T_1 \times \dots \times T_n \rightarrow U_1 \times \dots \times U_m$,
    \begin{align*}
      (\alpha) : ((C, t_1), \dots, (C,t_n)) &\mapsto ((C, \alpha(t_1, \dots, t_n)[1], \dots, \alpha(t_1, \dots, t_n)[m])) \\
      \text{anything else} &\mapsto \bot
    \end{align*}
  \end{defi}

  There is a functor $\Pi : \mathcal{T} \rightarrow \mathbf{Set}$, defined on objects by
  $\Pi([n_1, \dots, n_p]) = (\Theta \times 2^\Phi + \{ \bot \})^{n_1} \sqcup \dots \sqcup (\Theta \times 2^\Phi + \{ \bot \})^{n_p}$.
  Given a morphism $t : n \rightarrow p$, we define $\Pi(t)(\inj_i(x))$ as follows.
  If $x$ contains any $\bot$ or if the contexts in it are not all equal, then $\Pi(t)(\inj_i(x)) = \inj_1((\bot, \dots, \bot))$.\footnote{This is possible because we have assumed that codomains of
  morphisms in $\mathcal{T}$ are nonempty except for the identity on the monoidal unit.}
  Otherwise, as the truth tables $(t_{i,j})_j$ form a partition, there is a single case $(C,y)$ in all of them such
  that the associated CFF is true in the common context $C$. Let $j$ be the output index of its truth table: we set $\Pi(t)(\inj_i(x)) = \inj_j(y[x])$.
  One can check that this defines a monoidal functor.

  \begin{lemma} \label{lemma:chi-faithful}
    The functor $\Pi$ is faithful.
  \end{lemma}

  \begin{nestedproof}
    For simplicity, let us concentrate on the case of morphisms $t, t' : [n] \rightarrow [p]$:
    this is the only case that is actually needed to prove the completeness theorem, and the general
    case is similar.
    If $\Pi(t) = \Pi(t')$, then consider $t \otimes t' : n \rightarrow 2p$. For
    each case $(f,u,u') \in t \otimes t'$, with $f$ a CFF and $u, u'$ tuples of terms,
    $(f,u) = \Pi(t)(f,x_1, \dots, x_n) = \Pi(t')(f,x_1, \dots, x_n) = (f,u')$, so $u = u'$.
    Therefore $t \sim t'$.
  \end{nestedproof}

  Finally, combining Lemma~\ref{lemma:p-faithful} and Lemma~\ref{lemma:chi-faithful}, we obtain
  that $\Pi \circ P : \mathcal{F} \rightarrow \mathbf{Set}$ is faithful. But in fact $\Pi \circ P = S^*$, the functor arising from the valuation $S$. So, if two diagrams $d, d' \in \mathcal{F}$ give equal interpretations under any valuation $V$, then it is in particular the case for $V = S$, and by faithfulness of $S^*$, $d = d'$.
\end{proof}
We conjecture that this result generalizes to arbitrary morphisms in $\mathcal{F}$, with multiple input and output tables. However, all OpenRefine workflows have one input and one output table, so the theorem already covers these.

\section{Conclusion}

We have presented a complete axiomatization of the core data model of OpenRefine. This
gives a diagrammatic representation for workflows and an algorithm to determine if two
workflows are equivalent up to these axioms.

As future work, this visualization suggested by the categorical model
could be implemented in the tool itself. This would make workflows
easier to inspect, share and re-arrange. This representation could
also be the basis of a more profound overhaul of the implementation of
the data model, which would make workflow execution more scalable. The
axiomatization could also be extended to account for algebraic
equations involving the operations, although it seems hard to preserve
completeness and decidability for non-trivial equational theories.
Finally, the model could be extended to account for a larger class of
operations, for instance order-dependent ones such as sorting, or
operations which are not applied row-wise (using OpenRefine's
\emph{record} mode).

\section{Acknowledgements}

We thank David Reutter, Jamie Vicary and the anonymous reviewers for their helpful feedback on the project. The author is supported by an EPSRC Studentship.

\bibliographystyle{eptcs}
\bibliography{zotero}

\end{document}